\documentstyle[12pt]{article}
\begin{document}
\def\be{\begin{equation}}
\def\ee{\end{equation}}
\def\bea{\begin{eqnarray}}
\def\eea{\end{eqnarray}}
\def\non{\nonumber}
\def\ttgz{$t\overline{t}(\gamma ,Z)$~}
\def\cp{CP}
\def\g{\gamma}
\begin{flushright}
PRL-TH-97/24~~\\
hep-ph/9708332
\end{flushright}
\vskip 1cm
\begin{center}
{\Large \bf

CP-violating dipole form factors \mbox{of the top quark} and \mbox{tau
lepton} \mbox{in scalar leptoquark models}

}
\vskip .5cm
P. Poulose$^a$ and Saurabh D. Rindani$^{a,b}$
\vskip .25cm
$^a${\it Theory Group, Physical Research Laboratory\\
Navrangpura, Ahmedabad 380 009, India}\\
\vskip .25cm
$^b${\it Instituto de F\'\i sica Corpuscular - CSIC \\
Departament de F\'\i sica Te\` orica \\
Universitat de Val\` encia, 46100 Burjassot (Val\` encia), Spain}
\vskip 2cm
{\bf Abstract}
\end{center}
\vskip .25cm

We calculate the $CP$-violating electric and weak dipole form factors
of the top quark and the tau lepton in models with scalar leptoquarks coupling
only to the third generation of quarks and leptons. We obtain numerical
values of the real and imaginary parts of these form factors at various 
energies for different values of leptoquark
masses and couplings. The existing limits on the tau electric and weak
dipole form factors allows us to put a limit on the masses and couplings of
such leptoquarks and therefore on the top electric and weak dipole form factors.
We also discuss constraints on the form factors coming from indirect limits on
leptoquark masses and couplings deduced from LEP results on $Z$ properties.
\vskip .5cm

\noindent PACS Nos. 11.30.Er, 13.40.Em, 14.60.Fg, 14.65.Ha

\noindent Keywords: CP violation, scalar leptoquarks, dipole moments, top
quark, tau lepton

\newpage
\section { Introduction}

The standard model (SM) of electroweak interactions predicts that $CP$ 
violation outside the $K$- and $B$-meson systems would be unobservably
small. Thus, if any $CP$ violation is observed in the future outside of
these systems, it would be a signal of new physics.
In particular, the observation of electric dipole moments of 
elementary particles would signal new mechanisms of $CP$ violation lying  
beyond SM, since SM predicts 
unobservably small electric dipole moments. Likewise, their generalization
to $Z$ couplings (the so-called weak dipole moments), as well as to couplings
at nonzero momentum transfers, viz., dipole form factors (DFF), would also
serve as signals for $CP$ violation beyond SM.

Recent experiments at the Large Electron Positron Collider (LEP) at CERN
have established an upper bound on the weak dipole form factor of the tau lepton
at the $Z$ peak, by looking for $CP$-violating momentum correlations 
\cite{lepold,OPAL}. 
The latest result is from the OPAL collaboration \cite{OPAL}, who have obtained
the limits
$$ 
{\rm Re}\,d^Z_{\tau} < 5.6 \times 10^{-18} e {\rm cm} 
$$
$$
{\rm Im}\,d^Z_{\tau} < 1.5 \times 10^{-17} e {\rm cm}. 
$$
It is expected that better limits would be obtained in future experiments using
longitudinally polarized electrons \cite{Anant}. For example, it was shown that
using certain $CP$-odd vector correlations, it would be possible to measure
an EDFF of $\tau$ of the order of $10^{-19}$ $e$ cm \cite{Anant}.   

Estimates have also been made of the senstivity to which various experiments 
at $e^+e^-$ \cite{top} as well as hadron colliders \cite{hadron}
might be able to measure top electric and weak dipole form factors.

Most models seem to predict values of electric dipole form factors (EDFF)
and weak dipole form factors (WDFF) an order of magnitude
below the observable level in the next generation of planned experiments.
In this situation, it is a worthwhile pursuit to look at other models which 
might predict large DFF's which could be tested in the near future.

In this paper we consider the possibility that models with relatively light
scalars with complex couplings to a third-generation lepton-quark pair can 
give rise to 
EDFF and WDFF of third generation fermions, viz., the top quark and the tau
lepton, at a significant level. This possibility for the tau lepton 
has been considered recently
by Mahanta \cite{mahanta}  and Bernreuther {\it et al.} \cite{bernlq}. While
we confirm some results for tau of \cite{mahanta} and \cite{bernlq}, 
we also obtain
new results for the CP-violating form factors of the top quark, and treat 
the tau and top cases in a concerted manner. We have also studied the
implications of the existing experimental bounds from LEP on the weak dipole
form factor of $\tau$ at the $Z$ resonance, and on the masses and couplings of
leptoquarks from loop effects contributing to the $Z$ partial widths.

A large number of extensions of SM predict
the existence of colour triplet particles carrying simultaneously lepton 
and baryon number, called leptoquarks. These models
include grand unified theories, technicolour models, 
superstring inspired models and composite models. 
Without reference to specific models, the masses and couplings of 
leptoquarks can be constrained using low-energy experiments. These experiments
test predictions of leptoquark interactions for atomic parity violation, meson
decay, flavour-changing neutral currents and meson-antimeson mixing.

There have been several direct searches for leptoquarks at high energy
accelerators. At the Large Electron-Positron collider (LEP) at CERN,
a lower bound of 45-73 GeV for the mass of leptoquarks was put \cite{LEPbound}.
The limit coming from $p\overline{p}$ colliders is 175 GeV from D0 \cite{D0}
and 131-133 GeV from CDF \cite{CDF} 
on the mass of a scalar leptoquark decaying into an electron-jet pair.
On the mass of the third generation scalar leptoquark decaying into $b\tau$, 
CDF has given a bound of 99 GeV \cite{CDF} and a bound of 80 GeV was obtained
by D0 for such a leptoquark decaying into $b\nu_{\tau}$ \cite{D0}.
The excess of large-$Q^2$ events reported in $e^+p$ collisions at HERA 
\cite{HERA} have been interpreted as due to leptoquark production \cite{HERAlq},
the mass of the leptoquark being around 200 GeV.
The earlier limit coming from HERA is dependendent
on the leptoquark type and couplings, and the lower bound is between 92 and
184 GeV \cite{HERAlqold}.
Bounds possible at future $pp$, $ep$, $e^+e^-$, 
$e\gamma$ and $\gamma\gamma$ experiments has been a topic of serious study.

Indirect bounds on masses and couplings can be obtained from the results 
of low-energy experiments \cite{low-energy}. However, these constraints are 
strong only for leptoquarks that couple to quarks and leptons of the first 
and second generations.
 
While there are strong constraints on masses of leptoquarks
which also couple to pairs of quarks, thus violating baryon number as well 
as lepton number, the constraints on the couplings and masses of leptoquarks
which do not couple to two quarks are weaker. Moreover, these constraints
are strongest for the first and second generations, and considerably weaker
for leptoquarks coupling only to the third generation of quarks and leptons.

Strong constraints on leptoquarks which couple to leptons and quarks of the
third generation have been obtained from their contributions to the radiative
corrections to $Z$ properties \cite{Bhatta,Mizui,Eboli}. 
The authors of \cite{Bhatta}
have studied vertex corrections to the leptonic partial widths of the $Z$ 
induced by leptoquark loops and obtained stringent constraints on leptoquark
masses and couplings. The authors of \cite{Mizui,Eboli} 
performed a global fit to
the LEP data including contributions from a scalar leptoquark loop. They also
arrive at stringent constraints on leptoquark masses and couplings.

Earlier work on CP violation in leptoquark models can be found in
\cite{Hall,Geng}.

In this paper, we calculate the EDFF and WDFF for the third generation fermions,
viz., the top quark and the tau lepton, in scalar leptoquark models within 
the context
of an $SU(2)_L\times U(1)\times SU(3)_c$ gauge theory, where the leptoquarks
couple to only the third generation fermions. We then use the 
indirect limits obtained from LEP on the masses and couplings of the leptoquark
to investigate what are the possible values of the EDFF and WDFF in these
models consistent with these limits. 

Briefly, our results are as follows. The present experimental limits 
on $\tau$ EDFF and
WDFF do not put stringent constraints on masses and couplings of leptoquarks. 
Consequently, these results are consistent with a top EDFF of the order of
$10^{-19}\,e$ cm, and a top WDFF of the order of $10^{-20}\,e$ cm. The indirect
LEP limits, however, constrain these form factors to be at least three orders
of magnitude lower. They also give limits on tau form factors about two orders
of magnitude lower than the direct experimental limits.

We describe in the next section couplings of scalar leptoquarks with various
transformation properties under $SU(2)_L\times U(1)\times SU(3)_c$. In 
Section 3,
we give expressions for the imaginary parts of the EDFF's and WDFF's of
$\tau$ and $t$ arising from these leptoquark couplings, and write dispersion 
relations  for obtaining the corresponding real parts. In Section 4 we present
numerical results and the last section (Section 5) contains our conclusions.

\section{Scalar leptoquark couplings}
\label{sec:lqcouplings}

We will describe in this section leptoquark couplings in an 
$SU(2)_L\times U(1)\times SU(3)_c$ gauge theory, assuming that baryon-number 
violating couplings to diquarks are somehow forbidden, as required by strong
bounds on proton-decay searches. In that case, the only possible 
leptoquark representations which could have couplings to the 
standard-model representations of quarks and leptons are as shown in 
Table~\ref{tab:lqQn}, together with their quantum numbers \cite{Lqlag}.

The most general Lagrangian containing all possible forms of couplings of 
scalar leptoquarks to a lepton and quark pair  is given by
\be
{\cal L}_{eff}={\cal L}_{F=2}+{\cal L}_{F=0},\label{eq:lqleff}
\ee
where
\be
{\cal L}_{F=0}=h_{2L}\,\overline{u_R}\,R^T_2\,
i\tau_2\,l_L + h_{2R}\,\overline{q_L}\,e_R\,
R_2 + \tilde{h}_{2L}\,\overline{d_R}\,\tilde{R}^T_2\,i\tau_2\,l_L + h.c.,
\label{eq:lqlag1}
\ee
and
\be
{\cal L}_{F=2}=(g_{1L}\,\overline{q^c_R}\,i\tau_2\, l_L + g_{1R}\,
\overline{u^c_L}\,e_R)\,S_1 + 
\tilde{g}_{1R}\,\overline{d^c_L}\,e_R\,\tilde{S}_1 +
g_{3L}\,\overline{q^c_R}\,i
\tau_2\vec{\tau}\,l_L\cdot\vec{S_3} +  h.c., \label{eq:lqlag2}
\ee

The two pieces correspond to the fermion number $F=0$ for the
leptoquarks $R_2$ and $\tilde{R_2}$, and $F=-2$ for $S_1$, $\tilde{S_1}$
and $S_3$. Colour indices are suppressed in writing eqs.
(\ref{eq:lqlag1}) and (\ref{eq:lqlag2}). 

\begin{table}
\begin{center}
\begin{tabular}{||c|c|c|c|c||}
\hline
&$SU(2)_L$&$U(1)$&$SU(3)_c$&$Q$\\
\hline 
&&&&\\
$S_1$&1&$\frac{1}{3}$&$3^*$&$\frac{1}{3}$\\
$\tilde{S_1}$&1&$\frac{4}{3}$&$3^*$&$\frac{4}{3}$\\
$S_3$&3&$\frac{1}{3}$&$3^*$&$\frac{4}{3}$,$\frac{1}{3}$,$\frac{-2}{3}$\\
$R_2$&2&$\frac{7}{6}$&$3$&$\frac{5}{3}$,$\frac{2}{3}$\\
$\tilde{R_2}$&2&$\frac{1}{6}$&$3$&$\frac{2}{3}$,$\frac{-1}{3}$\\
&&&&\\
\hline 
\end{tabular}
\end{center}
\caption{\small 
$SU(2)_L$, $U(1)$, $SU(3)_c$ and electric charge 
assignments of the various leptoquarks}
\label{tab:lqQn}
\end{table}

Of the various couplings occurring in eqs.
(\ref{eq:lqlag1}) and (\ref{eq:lqlag2}),
those of $\tilde{S_1}$ and $\tilde{R_2}$ do not contribute to 
$\tau$ and $t$ DFF's in the limit of massless neutrinos
and $b$ quark, and so we will not consider these.

The low-energy constraints arising from decays of pseudoscalar mesons are very
stringent, unless the leptoquark couplings to the light quarks are chiral. 
Hence many authors assume couplings to be either 
left handed or right handed. However, we are going to consider only the 
third generation leptoquarks
on whose couplings there are no strong limits from meson decays. So we need not assume
their couplings to be chiral. More importantly, we need both left- and 
right-handed couplings to be present for the EDFF and WDFF to be nonzero. 
If, however, the third generation leptoquarks mix substantially with those of
the first and second generations, the low-energy constraints would apply more
or less unchanged. We will therefore assume mixing to be absent.

While only one of the components of each leptoquark multiplets would contribute
to the form factors we calculate, it is important to note that the constraints
we will use on the leptoquark couplings and masses were derived assuming that
leptoquarks within a multiplet are degenerate \cite{Bhatta,Mizui,Eboli}. These
constraints were also derived under the assumption of chiral couplings for
leptoquarks. However, in the presence of both left-handed and right-handed
couplings, the constraints are expected to be somewhat more stringent.
We have, however, chosen the constraints on left-handed
couplings for our analysis, since these are more stringent.

The couplings of leptoquarks to a single $\gamma$ or $Z$ is given by
\begin{equation}
{\cal L}_{\gamma ,Z}= -ie \sum_i \phi_i^{\dagger}\stackrel{\leftrightarrow}
{\partial}_{\mu}\phi_i \left[ Q_iA^{\mu}-\frac{T_{3L}-Q_is^2_W}{s_Wc_W}Z^{
\mu}\right] \label{eq:lqlgz},
\end{equation}
where $\phi_i$ are the various scalar fields, $T_{3i}$ and $Q_i$ are the 
respective values of the third component of weak isospin and electric charge,
and $c_W=\cos \theta_W$, $s_W=\sin \theta_W$, $\theta_W$ being the weak 
mixing angle.

We use the couplings written down in eqs. (\ref{eq:lqlag1}), 
(\ref{eq:lqlag2}) and (\ref{eq:lqlgz}) to obtain expressions for 
the EDFF and WDFF of $\tau$ and $t$ at the one-loop level in the next section.

\section{Electric and weak dipole form factors}
\label{sec:lqdff}

Using the leptoquark couplings 
described in the last section we shall look at the one-loop 
corrections to the \ttgz vertices.  We use Cutkosky rules \cite{cutrule}
to calculate
the imaginary part of the DFF and from this the real part is obtained 
using a dispersion relation.

\subsection{EDFF and WDFF of the top quark}

\begin{figure}
\vskip 4cm
\includegraphics{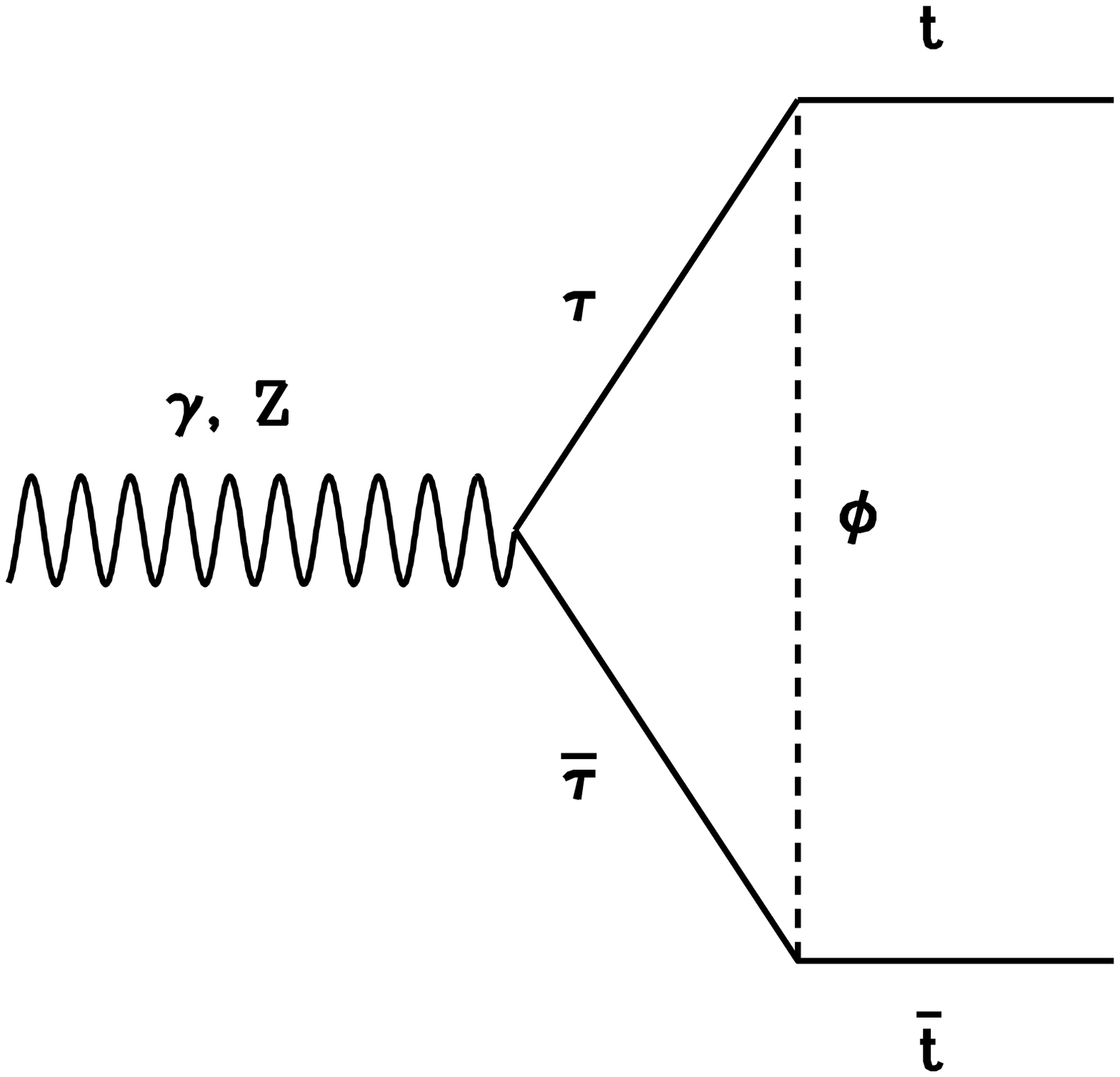}
\includegraphics{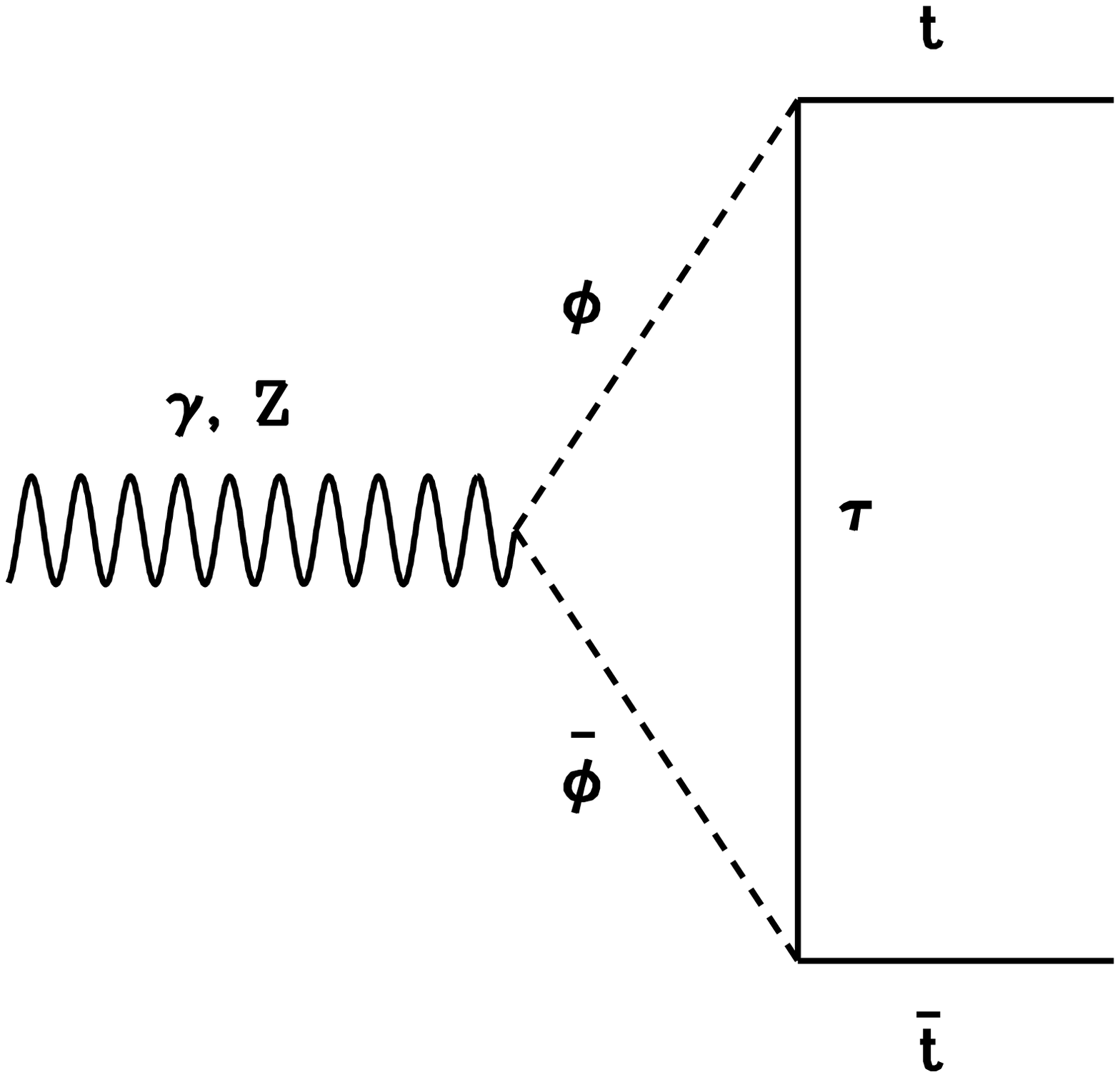}
\caption[dummy]
{\small Feynman diagrams corresponding to one-loop correction to the \ttgz 
vertex in the presence of a third generation leptoquark.}
\label{fig:feyndiag}
\end{figure}

At one-loop level the diagrams giving rise to the correction to the
\ttgz coupling due to the lepto\-quark are given 
in Fig.~\ref{fig:feyndiag}.
We use the symbol $\phi$ for the generic lepto\-quark.

The lepton could be $\nu _{\tau}$ instead of $\tau$ interacting
with a leptoquark of different $T_3$ and $Q$ values. However, as we shall
see, the contribution to the \cp-violating DFF's turns out to be
proportional to the mass of the virtual lepton because of chirality flip in
scalar couplings. In the limit of small $\nu _{\tau}$ mass, only the $\tau$
contribution is present.

We use Cutkosky rules to calculate the absorptive part of the process.  This
means considering intermediate particles to be on shell and hence the
conditions on the $\g/Z$ boson momentum $q$ for diagram in Fig.~1 are 
$q^2>4m^2_{\tau}$ and $q^2>4m^2_{\phi}$ in case of $\tau$ pair production
and leptoquark pair production respectively, where
$m_{\phi}$ is the leptoquark mass.  We calculate the vertex contribution
coming from these two diagrams.  

Using the couplings given in eqs. (\ref{eq:lqlag1}) and (\ref{eq:lqlag2})
we obtain the imaginary part of the top quark EDFF and WDFF as 
\bea
{\rm Im}\,d_t^{\gamma}(s)&=& 
	 \frac{eg_\phi^2}{4\pi s} m_{\tau}\,{\rm Im}\,(a^*b) 
	\left\{ - F_1^t (s) + Q F_2^t(s)\right\},\non \\
{\rm Im}\, d_t^Z(s) &=&  \frac{eg_\phi^2}{4\pi s \sin\theta_W \cos\theta_W}
 m_{\tau} {\rm Im}\,(a^*b) \non\\
&&\times \left\{\frac{1}{2} ( - \frac{1}{2} + 2 \sin^2 \theta_W) F_1^t (s)
	+ (T_3 - Q \sin^2\theta_W) F_2^t(s)\right\},
\eea
where
\bea
F_1^t (s) &=& \frac{1}{\beta_t^2} \left\{ \beta_{\tau} + \frac{1}{s\beta_t}
	(m_t^2 + m_{\phi}^2 -m_{\tau}^2) \right.\non  \\
&&\times \left.\ln\left[\frac{
	2\,(m_t^2 + m_{\tau}^2 - m_{\phi}^2) -s(1 - \beta_t \beta_{\tau})}
	{2\,(m_t^2 + m_{\tau}^2 - m_{\phi}^2)-s(1 + \beta_t \beta_{\tau})}
	\right] \right\}\theta (s-4 m_t^2),
\label{eq:lqimd1}
\eea
and
\bea
F_2^t (s) &=& \frac{1}{\beta_t^2} \left\{ \beta_{\phi} - \frac{1}{s\beta_t}
	(m_t^2 + m_{\phi}^2 -m_{\tau}^2- \frac{s}{2}) 
	\right.  \non \\
	&&\times \left.\ln\left[\frac{
	2\,(m_t^2 - m_{\tau}^2 + m_{\phi}^2) -s(1 - \beta_t \beta_{\phi})}
	{2\,(m_t^2 - m_{\tau}^2 + m_{\phi}^2)-s(1 + \beta_t \beta_{\phi})}
	\right] \right\}\theta (s-4 m_{\phi}^2).
\label{eq:lqimd2}
\eea

In the above equation, $T_3$ and $Q$ refer respectively to the third component
of isospin and charge of the leptoquark $\phi$, $g_\phi$ is 
the absolute value of the coupling constant, assuming $|g_L|=|g_R|$ 
(and $|h_{2L}|=|h_{2R}|$) occurring in eqs. (\ref{eq:lqlag1}) and 
(\ref{eq:lqlag2}),
and  $a$ and $b$ are phase factors of the corresponding vector and axial
vector couplings.
$\beta_t$, $\beta_{\tau}$ and $\beta_{\phi}$ refer to the velocities of
$t$, $\tau$ and $\phi$,
\be
\beta_{t,\tau,\phi} = \sqrt{1- \frac{4 m^2_{t,\tau,\phi}}{s}}.
\ee
The expression in  eq. (\ref{eq:lqimd1})
is valid for $s>4 m_t^2$. $F_1^t(s)$ for $2 m_{\tau}
<\sqrt{s}<2 m_t$ is given by the analytic continuation of eq.
(\ref{eq:lqimd1}): 
\bea
F_1^t (s) &=& \frac{1}{\beta_t^2} \left\{ \beta_{\tau} + \frac{2}{s\sqrt{
-\beta^2_t}}
	(m_t^2 + m_{\phi}^2 -m_{\tau}^2) \right.\non \\
&&\times \tan ^{-1} \left[\left. \frac{
\sqrt{-\beta^2_t}\,\beta_{\tau}\,s}
	{2\,(m_t^2 + m_{\tau}^2 - m_{\phi}^2) -s}\right] \right\}
	\theta (s-4 m_\tau^2)\theta (4 m_t^2-s)
\eea

The real parts of the form factors are obtained using an unsubtracted
dispersion relation
\be
{\rm Re}\,d_t^{\gamma ,Z} (s) = \frac{\rm P}{\pi} \int^{\infty}_{4m^2_{\tau}}
\frac{{\rm Im}\,d_t^{\gamma ,Z} (s')}{ s' -s}\,ds',
\ee
where $P$ denotes the principal part of the integral.

It can be seen that the dispersion integrals are convergent and do not need any
subtraction. This is to be expected since the dipole form factors, 
which correspond to dimension 5 operators, should be finite in a renormalizable
theory.

\subsection{EDFF and WDFF of the $\tau$ lepton}

The EDFF and WDFF of the tau lepton also arise in the same leptoquark 
theory from diagrams exactly 
analogous to the ones in Fig.~1, with the roles of $t$ and $\tau$
interchanged. Proceeding exactly as in the previous section, we obtain 
expressions for the imaginary parts of the tau lepton EDFF and WDFF, 
neglecting the $b$ mass.
\bea
{\rm Im}\, d_{\tau}^{\gamma}(s) &=&  
	\frac{3eg_\phi^2}{4\pi s} m_t {\rm Im}(a^*b)\left\{ \frac{2}{3}
 	F_1^{\tau} (s) - Q F_2^{\tau}(s)\right\},\non \\
{\rm Im}\,d_{\tau}^Z(s) &=&  \frac{3eg_\phi^2}{4\pi s \sin\theta_W \cos\theta_W}
 m_t {\rm Im}(a^*b) \non \\
&&\times \left\{\frac{1}{2} 
(  \frac{1}{2} - \frac{4}{3} \sin^2 \theta_W) F_1^{\tau}
 (s)
	- (T_3 - Q \sin^2\theta_W) F_2^{\tau}(s)\right\},
\eea
where
\bea
F_1^{\tau} (s)&=& \frac{1}{\beta_{\tau}^2} \left\{ \beta_t + \frac{1}{s\beta_{\tau}}
	(m_{\tau}^2 + m_{\phi}^2 -m_t^2)   \right. \non \\
	&&\times \ln \left[\left.\frac{
	 2\,(m_{\tau}^2 + m_t^2 - m_{\phi}^2) -s\,(1 - \beta_t \beta_{\tau})}
	{2\,(m_{\tau}^2 + m_t^2 - m_{\phi}^2) -s\,(1 + \beta_t \beta_{\tau})}
	\right] \right\}\theta (s-4 m_t^2),
\eea
and
\bea
F_2^{\tau} (s) &=& \frac{1}{\beta_{\tau}^2} 
	\left\{ \beta_{\phi} - \frac{1}{s\beta_{\tau}}
	(m_{\tau}^2 + m_{\phi}^2 -m_t^2- \frac{s}{2})  \right. \non \\
	&&\times \ln \left[\left.\frac{
       2\,(m_{\tau}^2 - m_t^2 + m_{\phi}^2) -s\,(1 - \beta_{\tau} \beta_{\phi})}
      {2\,(m_{\tau}^2 - m_t^2 + m_{\phi}^2) -s\,(1 + \beta_{\tau} \beta_{\phi})}
	\right] \right\}\theta (s-4 m_{\phi}^2).
\eea
Since both $m_t$ and $m_{\phi}$ are larger than $m_{\tau}$, there is no 
domain where an analytic continuation is needed.

As before, the real parts of the form factors are given by the unsubstracted
dispersion relations:
\be
{\rm Re}\,d_{\tau}^{\gamma ,Z} (s) = 
	\frac{\rm P}{\pi} \int^{\infty}_{4m^2_{\tau}}
\frac{{\rm Im}\,d_{\tau}^{\gamma ,Z} (s')}{ s' -s}\,ds'.
\ee

In the next section we will evaluate the real and imaginary parts of the $t$ 
and $\tau$ form factors numerically for different choices of masses and
couplings of the leptoquarks.
Using the experimental limits on the tau lepton DFF's we obtain bounds on
the masses and couplings of the leptoquarks. 

\section{Numerical results}

\begin{figure}
\vskip 5cm
\includegraphics{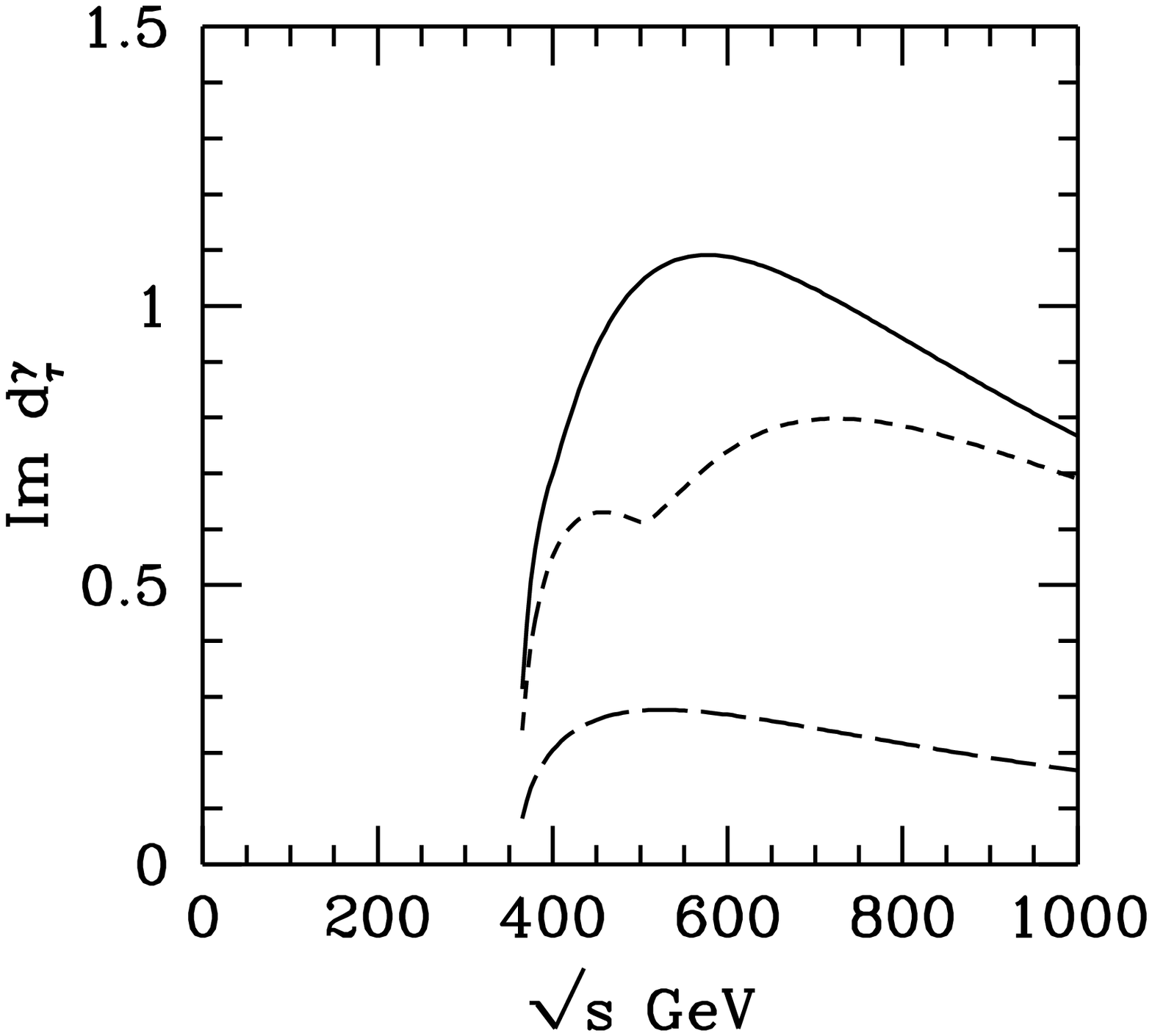}
\includegraphics{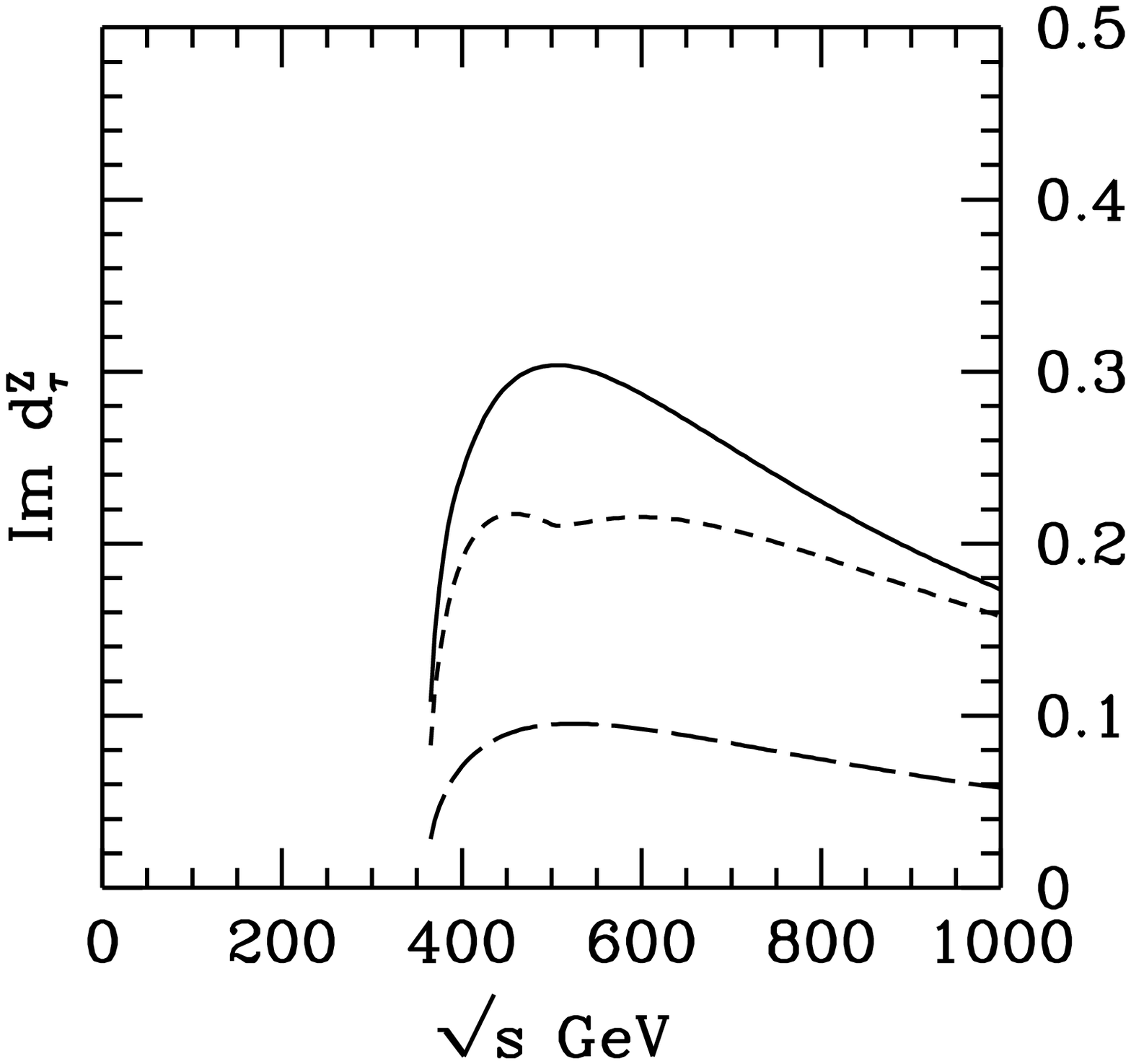}
\caption[dummy]
{\small Imaginary parts of the electric (figure on the left) and weak (figure on the
right)
dipole form factors of $\tau$ in units of $10^{-18}\;e$ cm
as functions of c.m. energy $\sqrt{s}$ for the model with leptoquark $R_2$. 
Solid, dashed and dash-dotted
lines correspond to leptoquark masses of 200 GeV, 250 GeV
and 500 GeV respectively. 
$g_\phi$ is chosen to be 1.
}
\label{fig:tauimgcme1}
\end{figure}

\begin{figure}
\vskip 5cm
\includegraphics{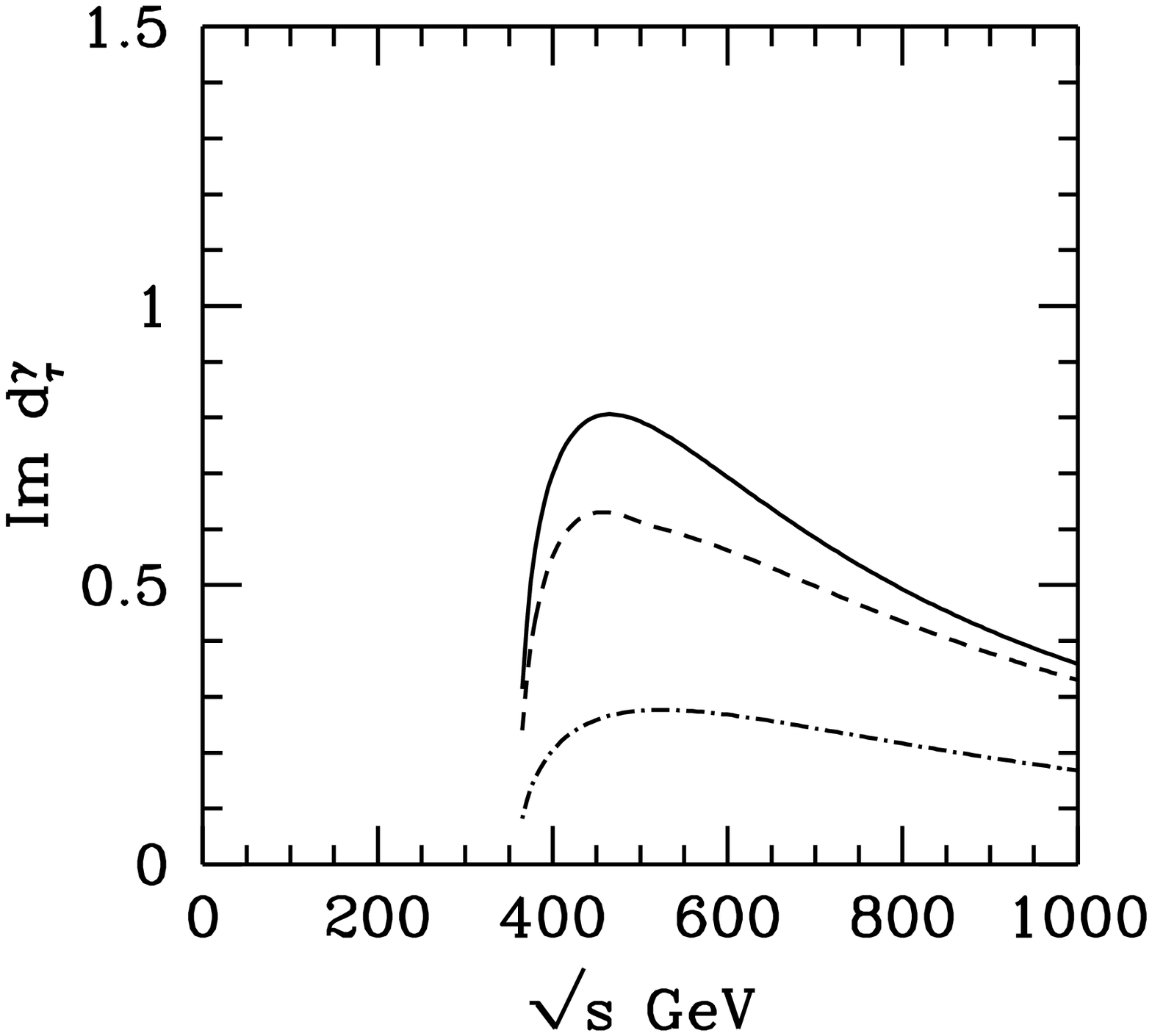}
\includegraphics{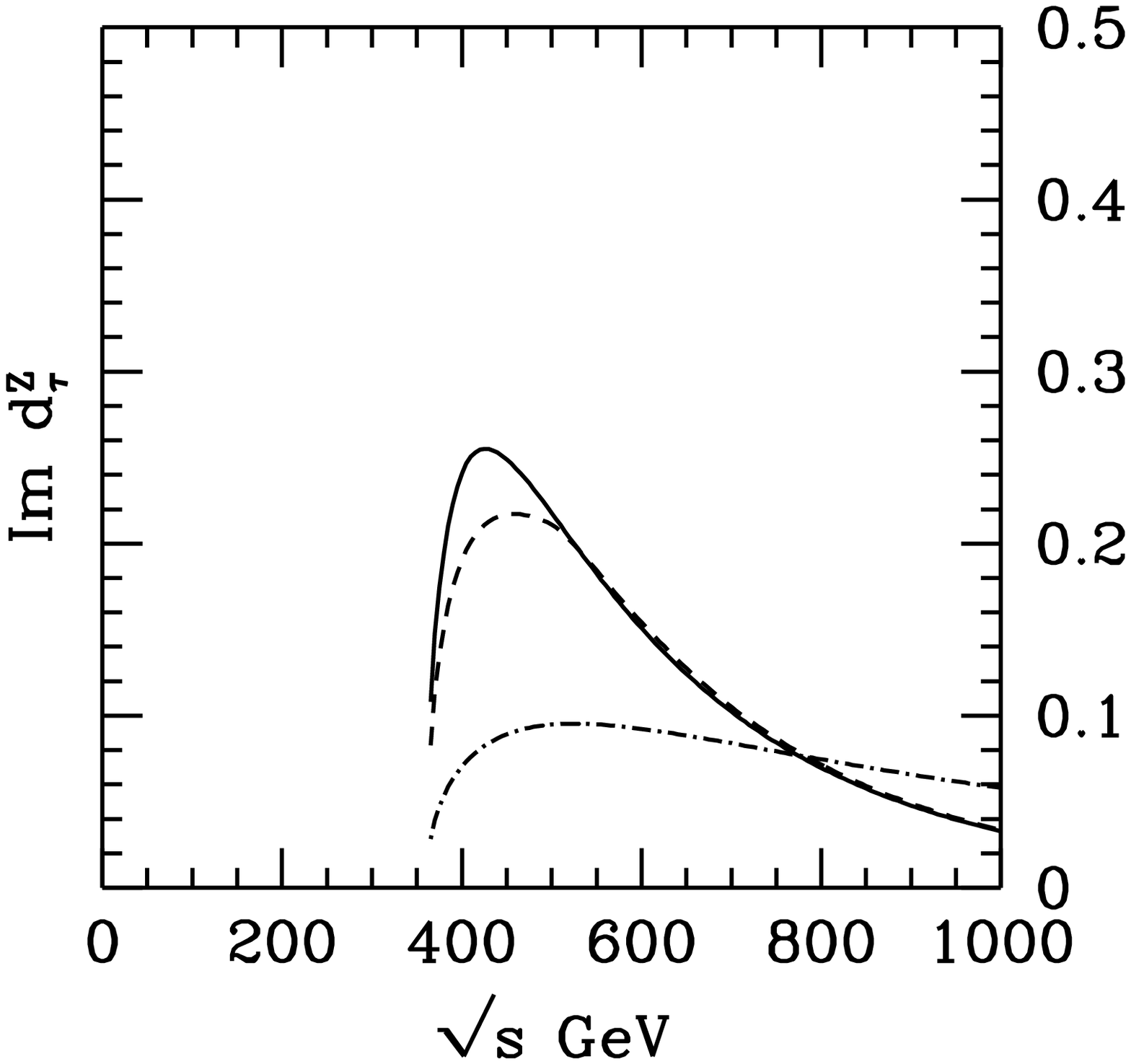}
\caption[dummy]
{\small Imaginary parts of the electric (figure on the left) and weak (figure on
the right)
dipole form factors of $\tau$ in units of $10^{-18}\;e$ cm
as functions of c.m. energy $\sqrt{s}$ for the model with leptoquark $S_1$. 
Solid, dashed and dash-dotted
lines correspond to leptoquark masses of 200 GeV, 250 GeV
and 500 GeV respectively. 
$g_\phi$ is chosen to be 1.
}
\label{fig:tauimgcme2}
\end{figure}

\begin{figure}
\vskip 4.5cm
\includegraphics{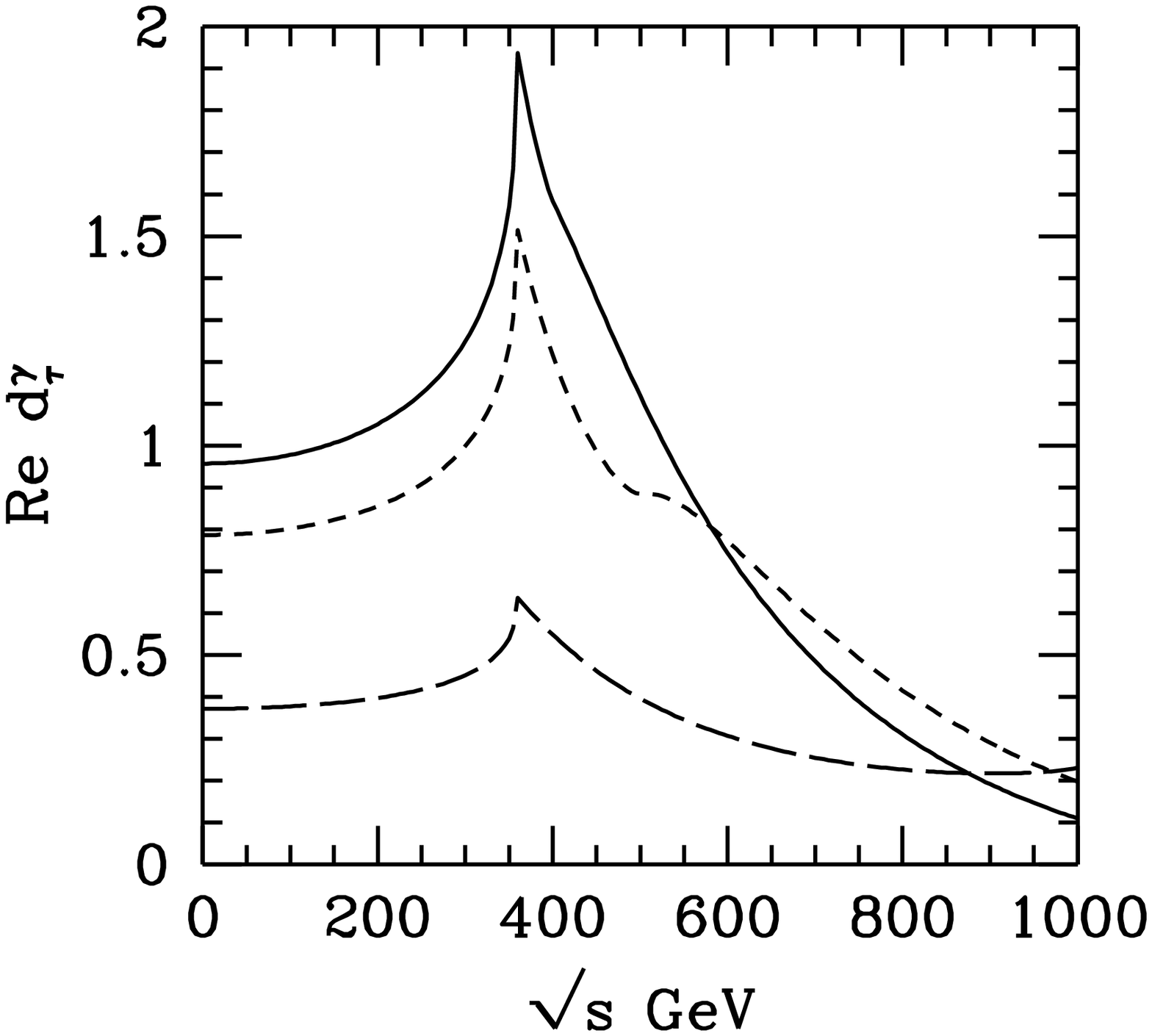}
\includegraphics{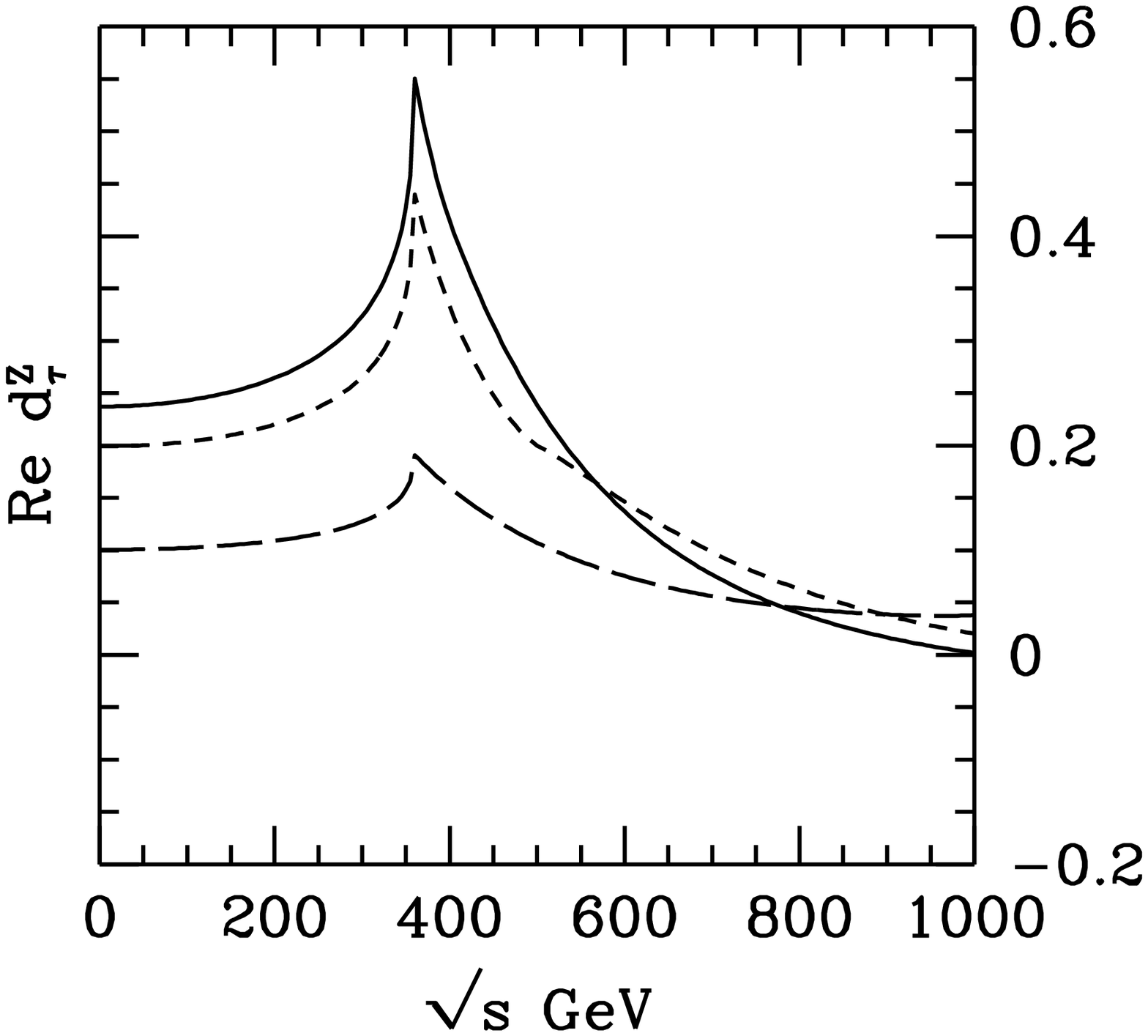}
\caption[dummy]
{\small Real parts of the electric (figure on the left) and weak (figure on the
right)
dipole form factors of $\tau$ in units of $10^{-18}\;e$ cm
as functions of c.m. energy $\sqrt{s}$ for the model with leptoquark $R_2$. 
Solid, dashed and dash-dotted
lines correspond to leptoquark masses of 200 GeV, 250 GeV
and 500 GeV respectively. 
$g_\phi$ is chosen to be 1.
}
\label{fig:taurealcme1}
\end{figure}

\begin{figure}
\vskip 5cm
\includegraphics{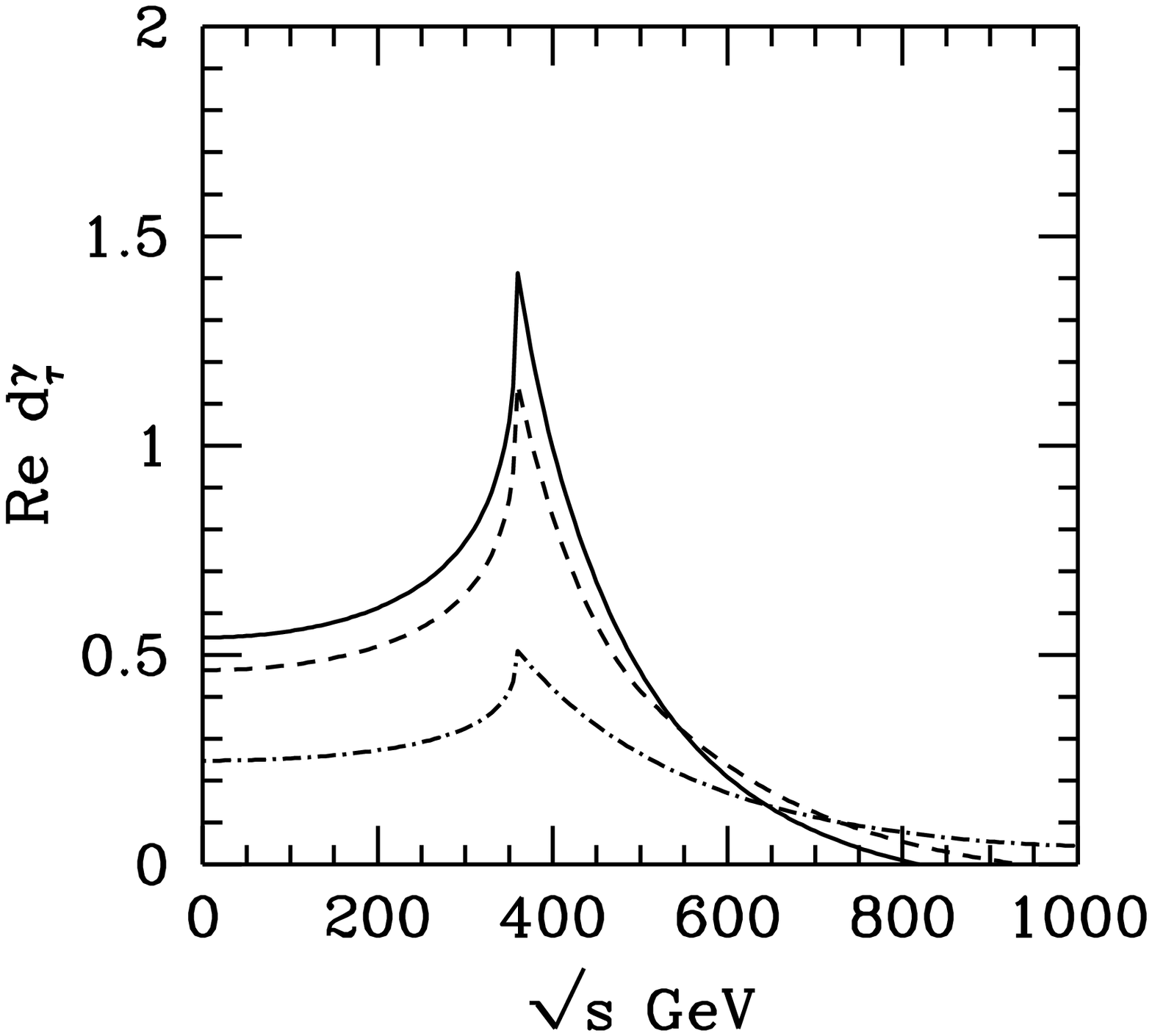}
\includegraphics{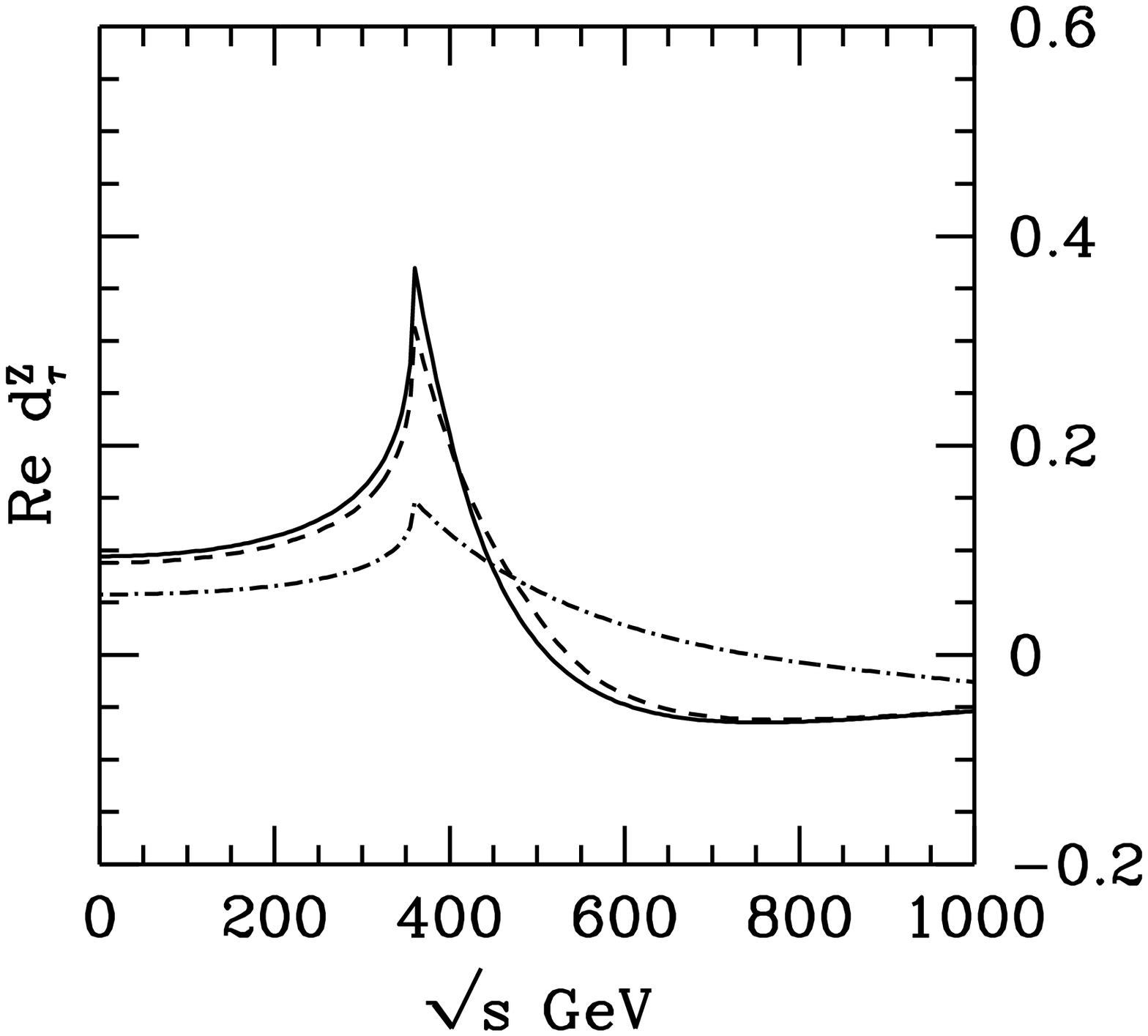}
\caption[dummy]
{\small Real parts of the electric (figure on the left) and weak (figure on 
the right)
dipole form factors of $\tau$ in units of $10^{-18}\;e$ cm
as functions of c.m. energy $\sqrt{s}$ for the model with leptoquark $S_1$. 
Solid, dashed and dash-dotted
lines correspond to leptoquark masses of 200 GeV, 250 GeV
and 500 GeV respectively. 
$g_\phi$ is chosen to be 1.
}
\label{fig:taurealcme2}
\end{figure}

We use here the expressions of the previous section to get numerical values for
the various form factors. While we have analytic expressions for the imaginary
parts, the real parts, given by the dispersion integrals, have been evaluated
by numerical integration.

To investigate how large the form factors can be,
consistent with LEP constraints, we have plotted 
the $\tau$ and $t$ form factors as functions of $\sqrt{s}$. 
We have chosen $m_t= 180$ GeV
and a maximal value Im $(a^*b)=1/2$.
We have considered three different leptoquark masses, viz., 200 GeV,  250
GeV and 500 GeV. $g_\phi$ is chosen to be 1.
In all cases dipole form factors are larger for higher leptoquark masses
upto around c.m. energy of 500 GeV.  At higher energies dependence on the
mass becomes weaker than that at lower energies.  Among the leptoquarks
belonging to the three representations considered, $R_2$
(see Table~\ref{tab:lqQn} for quantum numbers) gives the largest form
factors. 
The leptoquarks $S_3$ (isospin triplet) and $S_1$ (isospin singlet) give
the same values for DFF's except for the sign, and so we have presented only
the results for the $S_1$ model. 

Figs.~\ref{fig:tauimgcme1} and
\ref{fig:tauimgcme2} show the dependence of the imaginary parts of the
$\tau$ EDFF and WDFF on the c.m. energy. Similarly,
Figs.~\ref{fig:taurealcme1} and
\ref{fig:taurealcme2} show the dependence of the real parts.
Since the top quark mass is assumed to be 180 
GeV, there is a peak at 360 GeV. A similar, but not
so prominent, behaviour is seen at the leptoquark threshold.

\begin{figure}
\vskip 5.5cm
\includegraphics{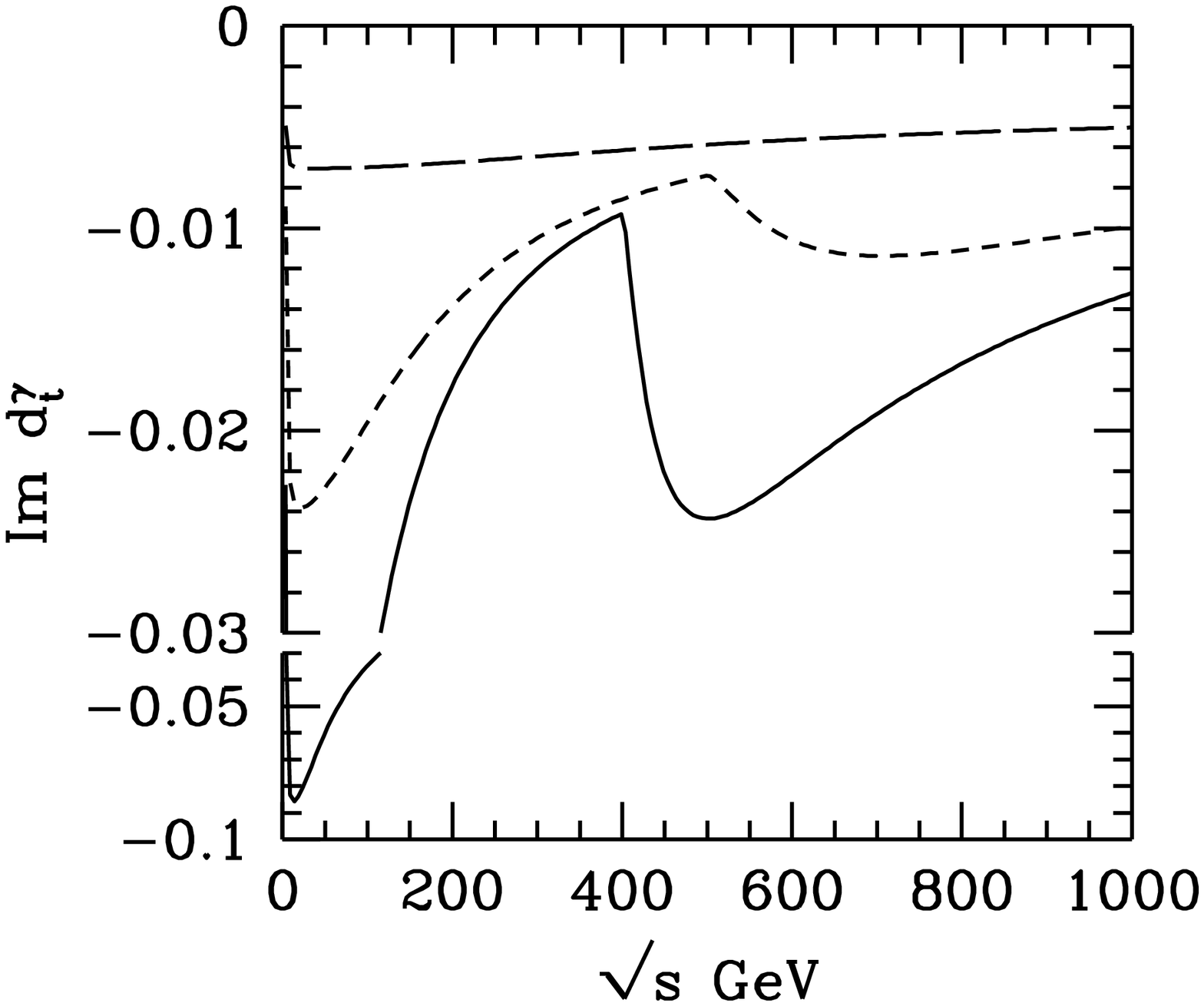}
\includegraphics{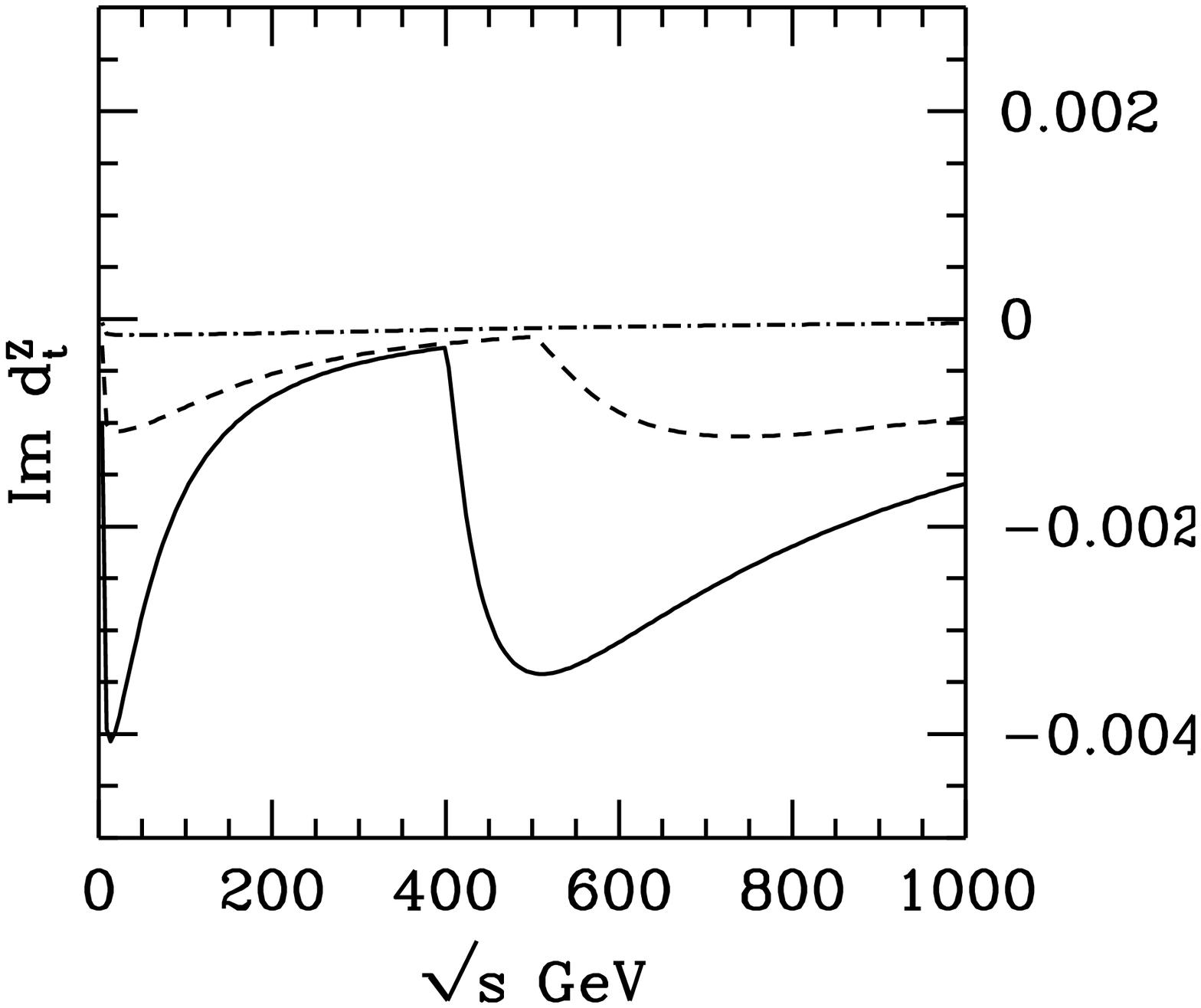}
\caption[dummy]
{\small Imaginary parts of the electric (figure on the left) and weak (figure on the
right)
dipole form factors of the top quark in units of $10^{-18}\;e$ cm
as functions of c.m. energy $\sqrt{s}$ for the model with leptoquark $R_2$. 
Solid, dashed and dash-dotted
lines correspond to leptoquark masses of 200 GeV, 250 GeV
and 500 GeV respectively. 
$g_\phi$ is chosen to be 1.
}
\label{fig:topimgcme1}
\end{figure}

\begin{figure}
\vskip 5cm
\includegraphics{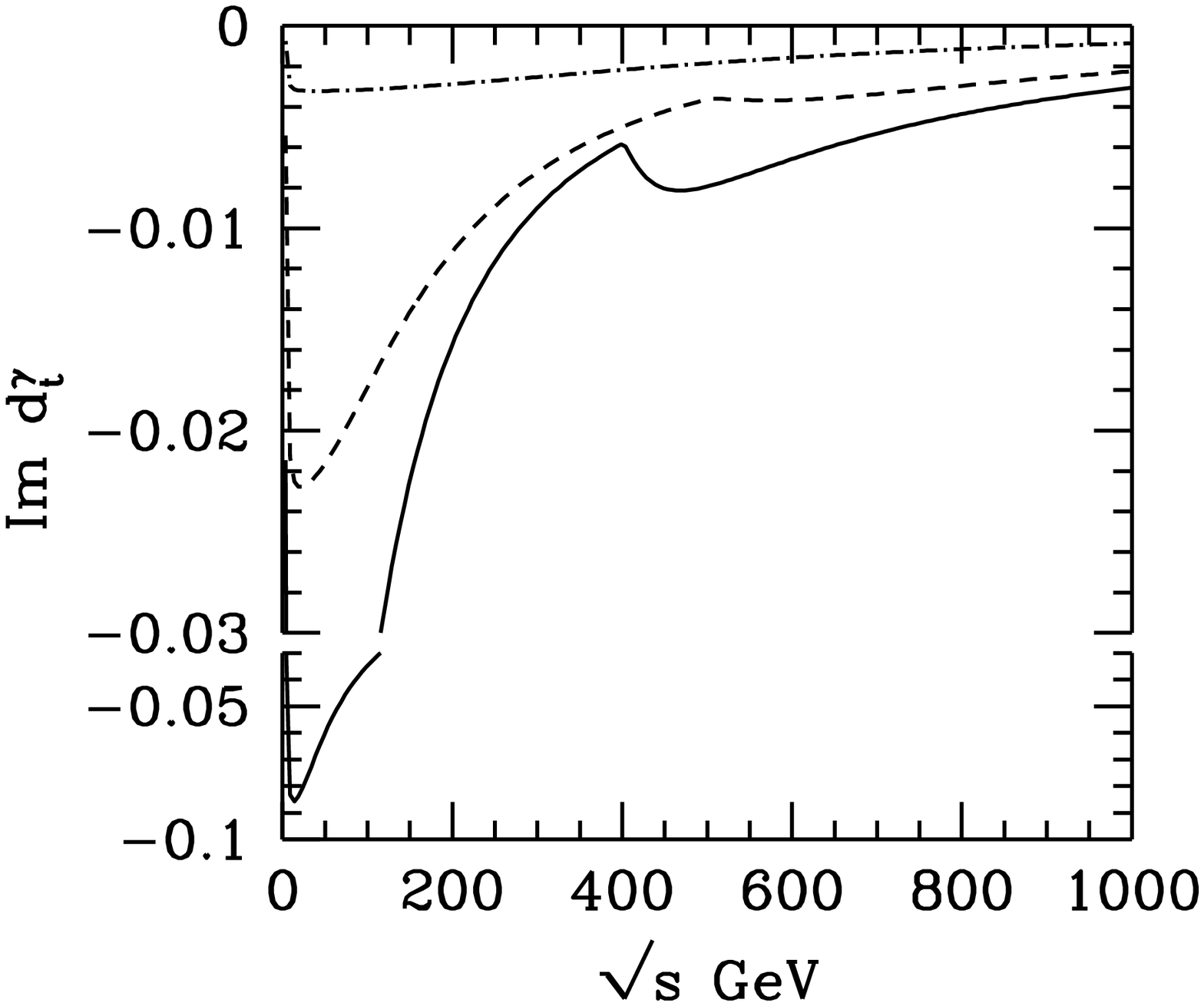}
\includegraphics{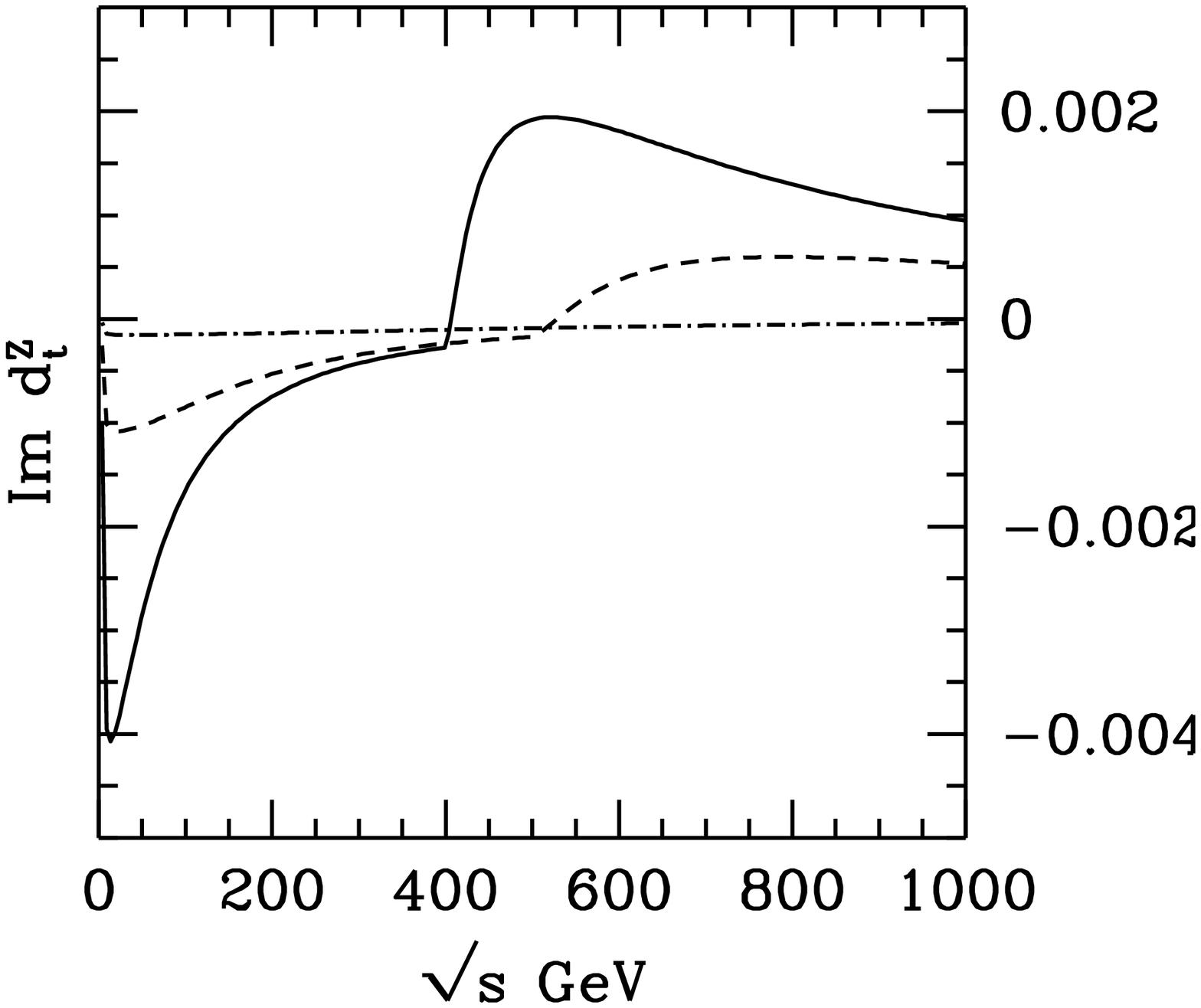}
\caption[dummy]
{\small Imaginary parts of the electric (figure on the left) and weak (figure on 
the right)
dipole form factors of the top quark in units of $10^{-18}\;e$ cm
as functions of c.m. energy $\sqrt{s}$ for the model with leptoquark $S_1$. 
Solid, dashed and dash-dotted
lines correspond to leptoquark masses of 200 GeV, 250 GeV
and 500 GeV respectively. 
$g_\phi$ is chosen to be 1.
}
\label{fig:topimgcme2}
\end{figure}

\begin{figure}
\vskip 5cm
\includegraphics{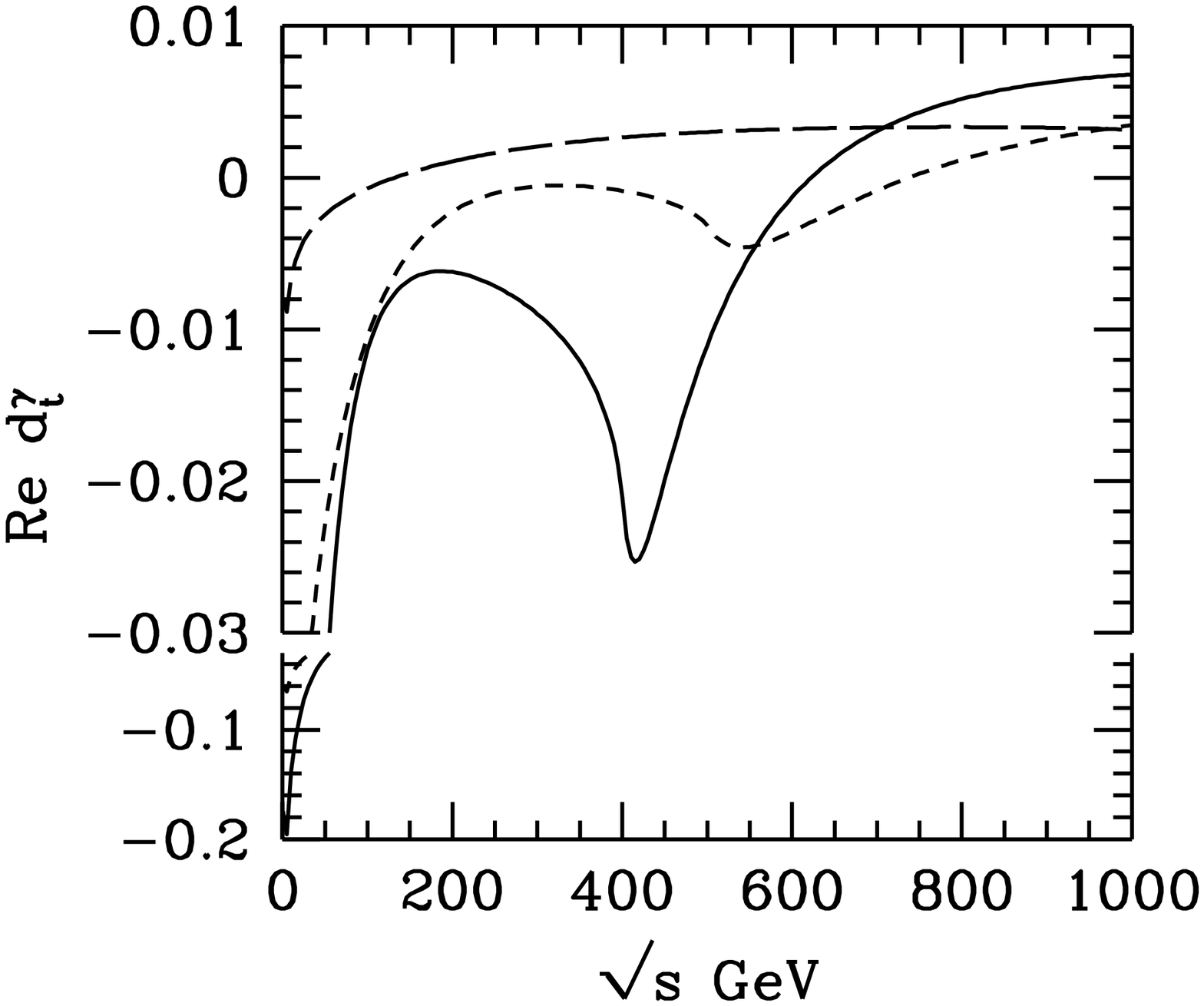}
\includegraphics{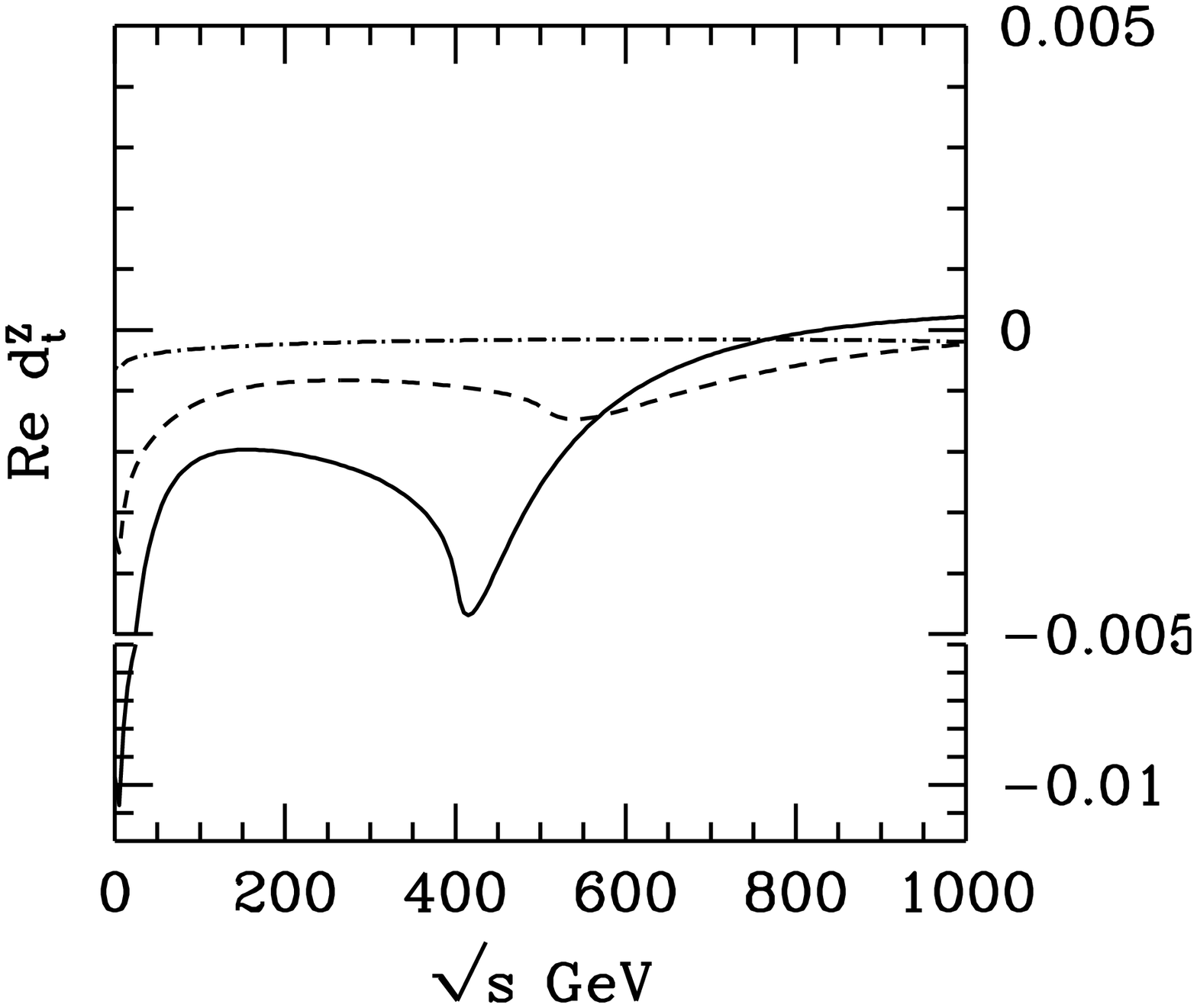}
\caption[dummy]
{\small Real parts of the electric (figure on the left) and weak (figure on the
right)
dipole form factors of the top quark in units of $10^{-18}\;e$ cm
as functions of c.m. energy $\sqrt{s}$ for the model with leptoquark $R_2$. 
Solid, dashed and dash-dotted
lines correspond to leptoquark masses of 200 GeV, 250 GeV
and 500 GeV respectively. 
$g_\phi$ is chosen to be 1.
}
\label{fig:toperealcme1}
\end{figure}

\begin{figure}
\vskip 5cm
\includegraphics{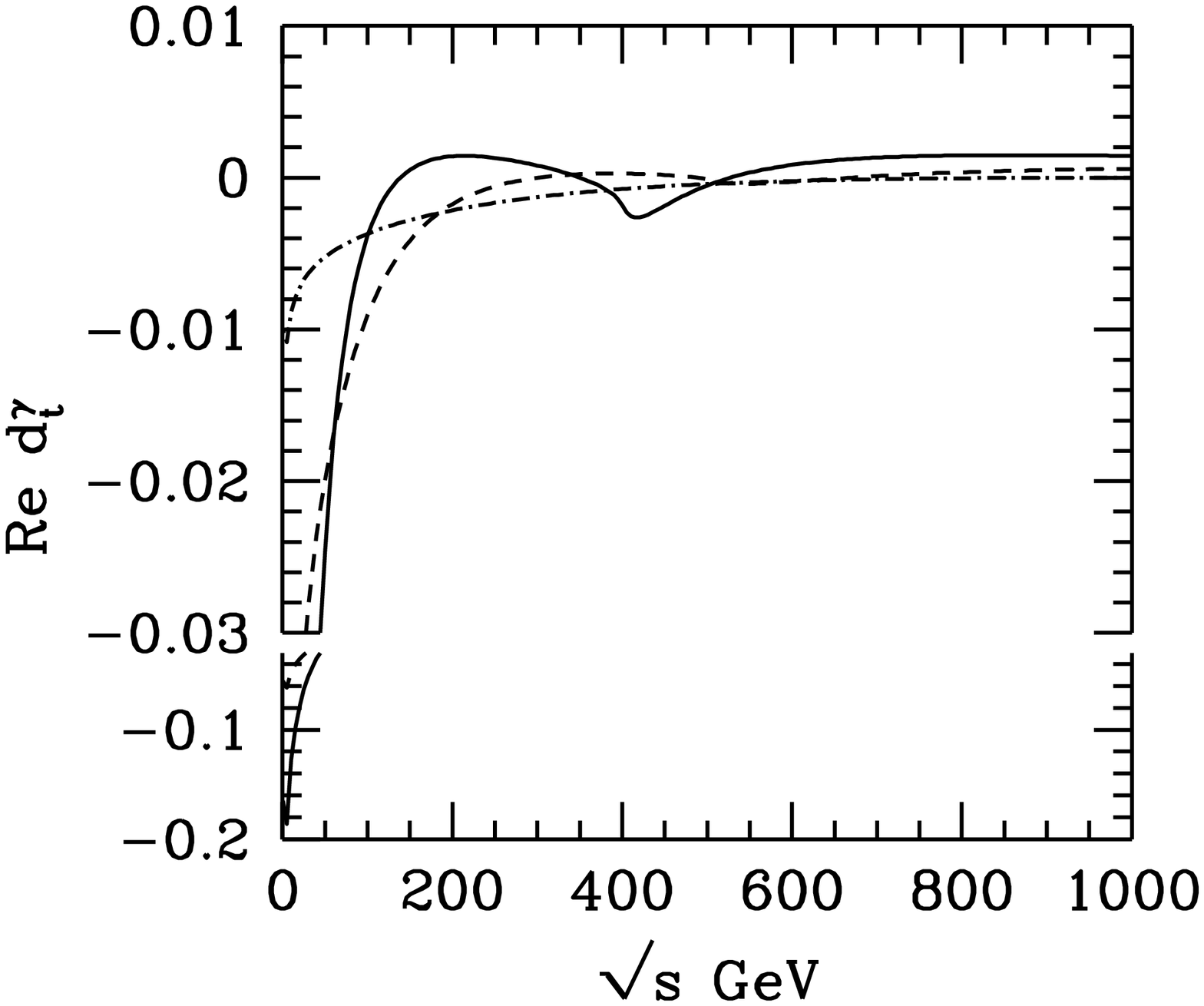}
\includegraphics{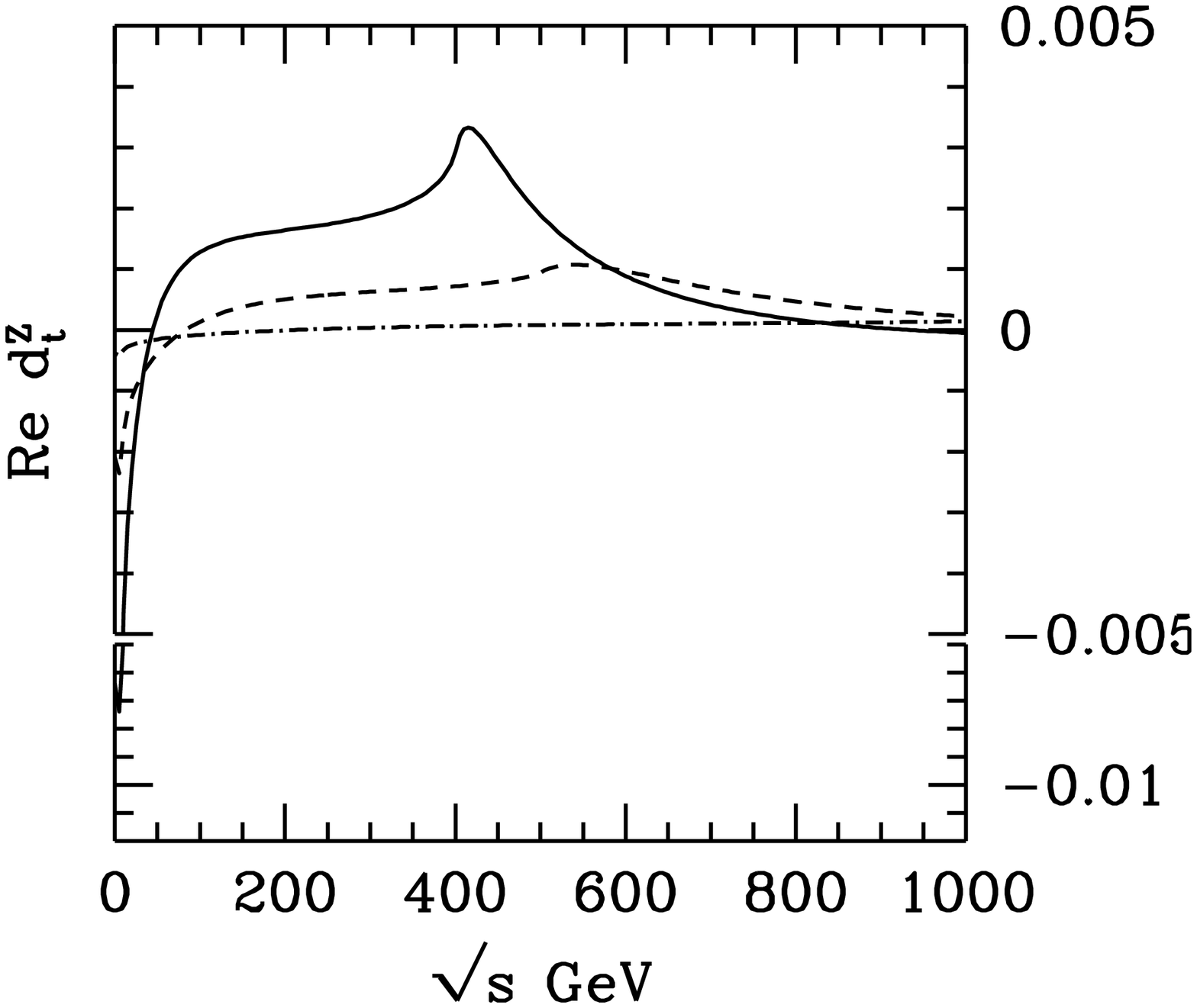}
\caption[dummy]
{\small Real parts of the electric (figure on the left) and weak (figure on 
the right)
dipole form factors of the top quark in units of $10^{-18}\;e$ cm
as functions of c.m. energy $\sqrt{s}$ for the model with leptoquark $S_1$. 
Solid, dashed and dash-dotted
lines correspond to leptoquark masses of 200 GeV, 250 GeV
and 500 GeV respectively. 
$g_\phi$ is chosen to be 1.
}
\label{fig:toperealcme2}
\end{figure}

Figs.~\ref{fig:topimgcme1} and \ref{fig:topimgcme2} show 
the variation of the imaginary part of the $t$ DFF's with c.m. energy, and 
figs.~\ref{fig:toperealcme1} and \ref{fig:toperealcme2} show the corresponding
curves for the real parts\footnote{ 
Note that in the case of the top quark,
the figures corresponding imaginary and real parts of EDFF, and 
real part of WDFF, are plotted at in two parts,
with different scales for the $y$ axis in the two parts.}.
The behaviour is similar to that in the tau case with peaks at $\tau$ and
leptoquark resonances.
In case of the top quark form factors, it is interesting to observe 
the difference
in signs, especially of the WDFF, between the $R_2$ and $S_1$ leptoquark
models.

It is seen from the curves that in general, the $R_2$ model gives larger values
of the form factors compared to the $S_1$ model. We therefore concentrate on
the $R_2$ model in what follows.

\begin{figure}
\vskip 5cm
\includegraphics{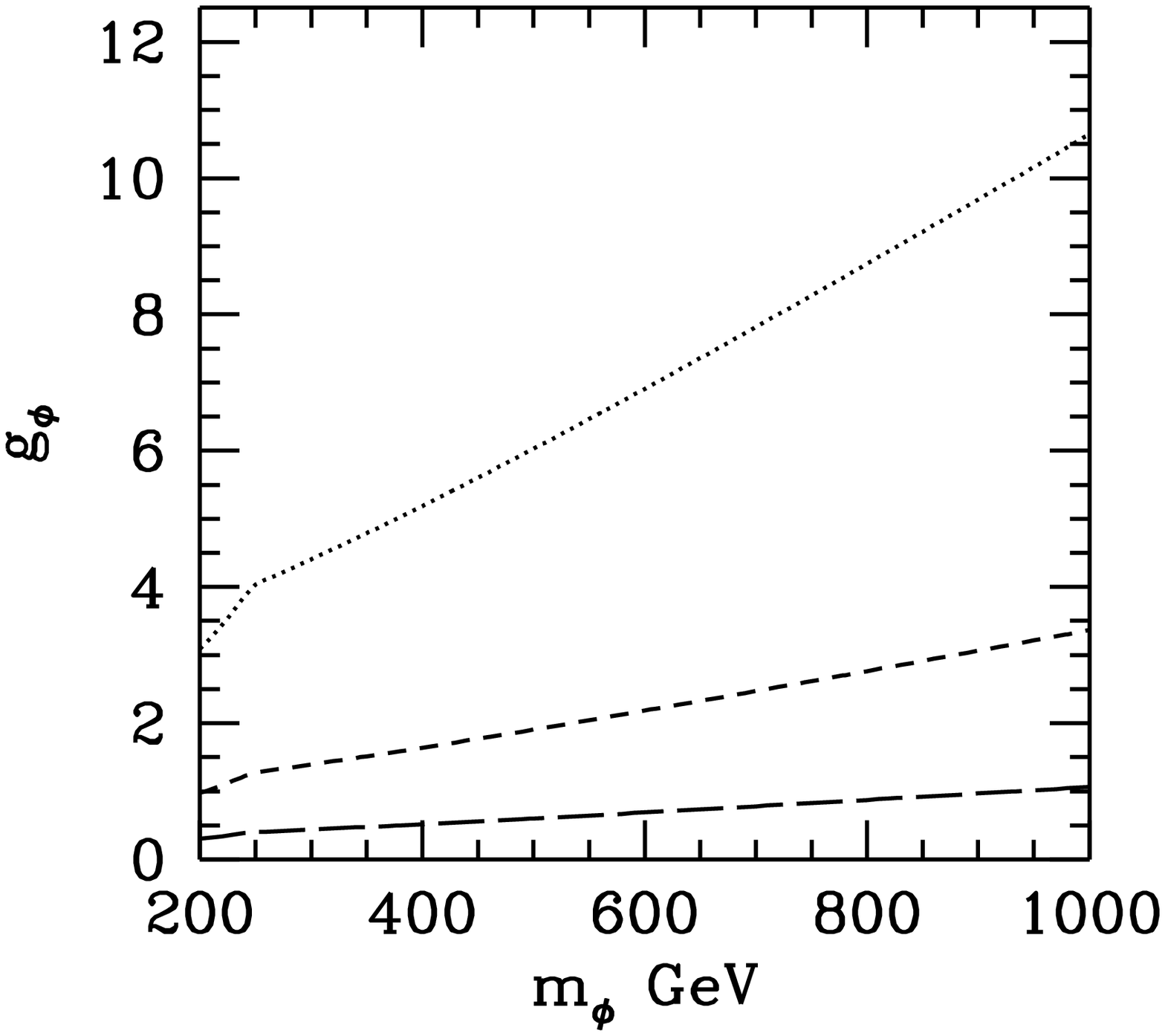}
\includegraphics{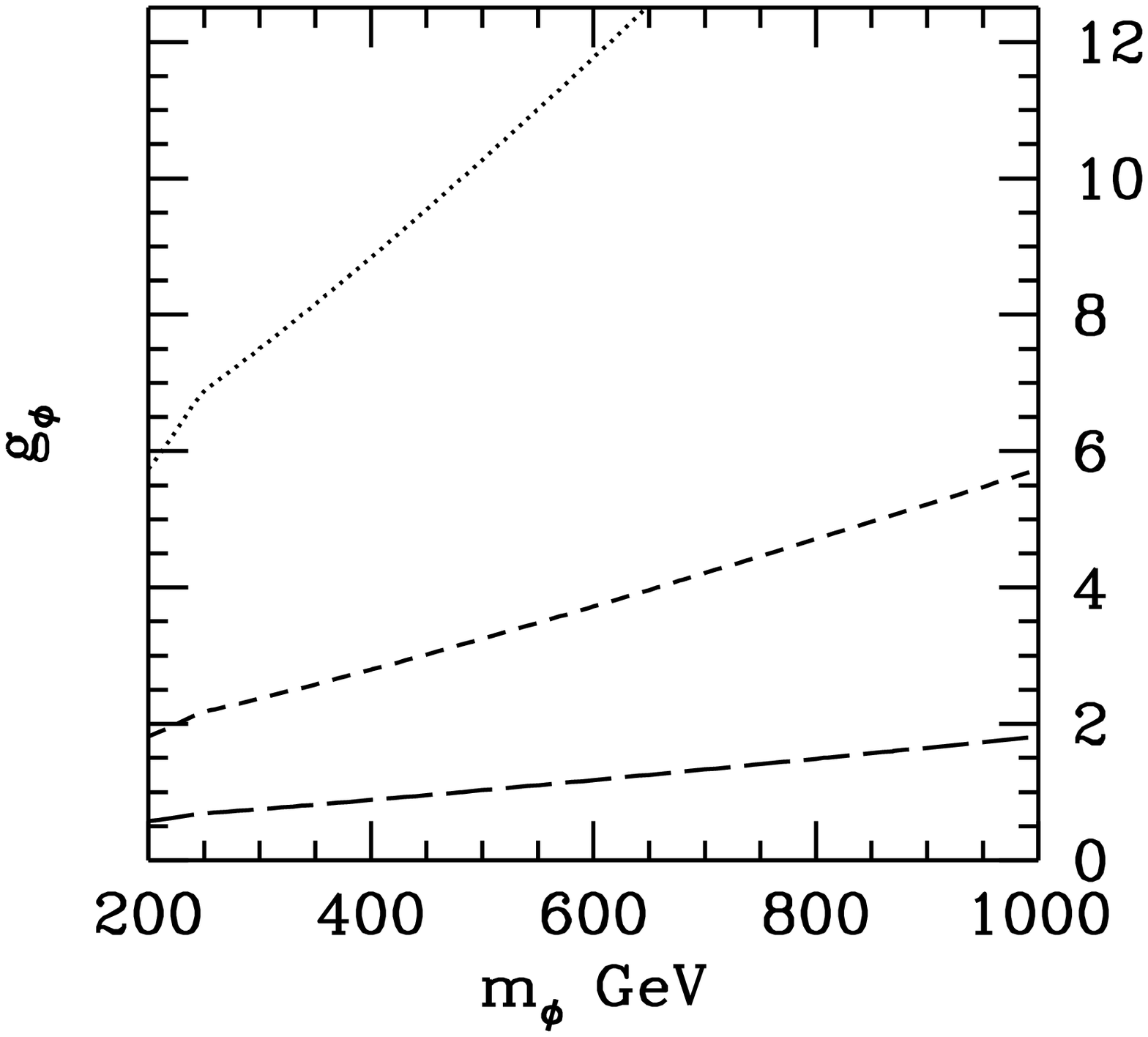}
\caption[dummy]{\small Contours in the mass-coupling plane for different values of 
imaginary part of EDFF (left) and WDFF (right) of $\tau$ for the $R_2$ model.
Dotted, dashed, long dashed and dash-dotted lines correspond to DFF values of 
$10^{-17}$, $10^{-18}$, $10^{-19}$ and $10^{-20}\;e$ cm respectively. 
A c.m. energy of 500 GeV is assumed.}
\label{fig:conttauimg}
\end{figure}

\begin{figure}
\vskip 5.5cm
\includegraphics{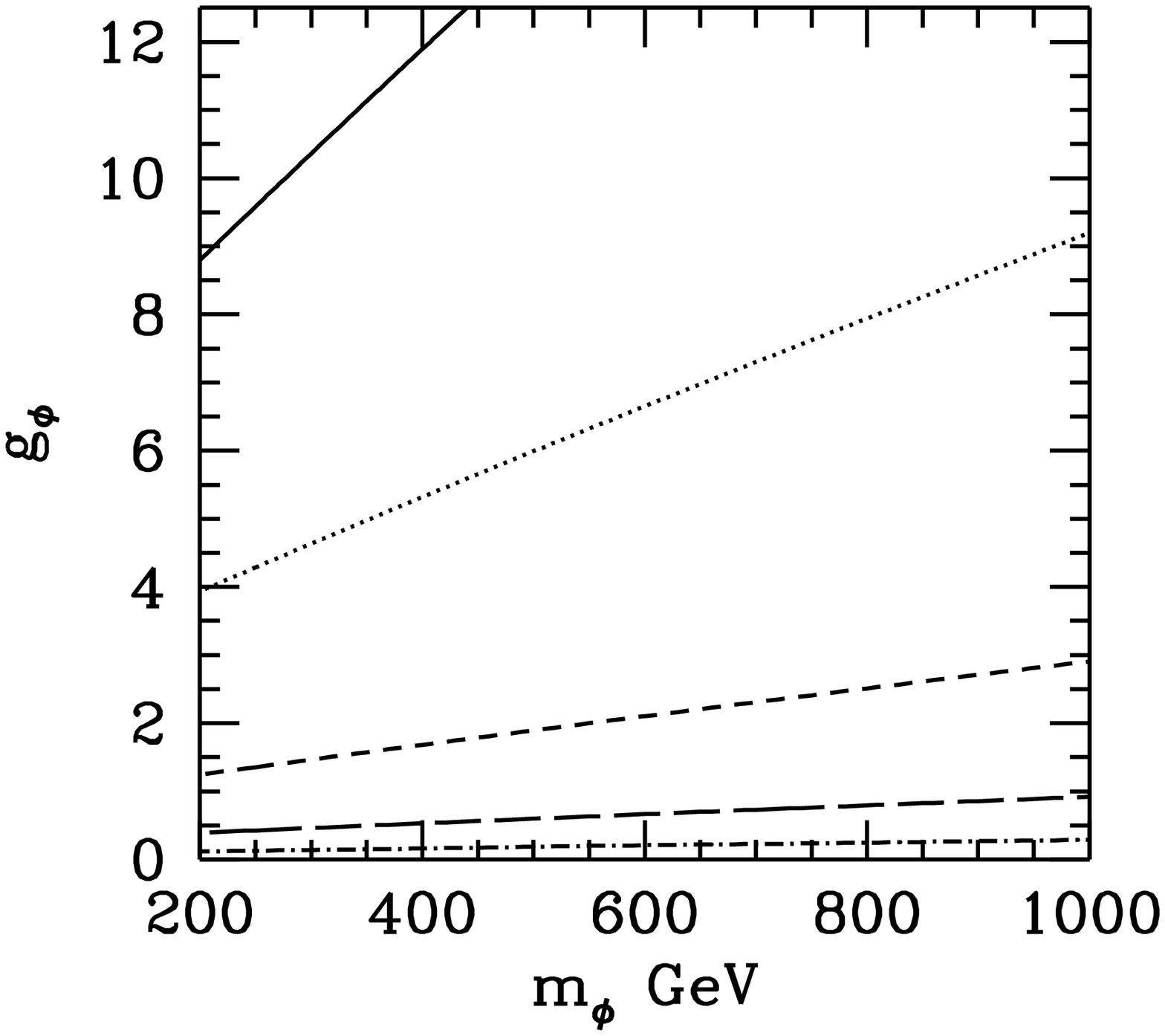}
\includegraphics{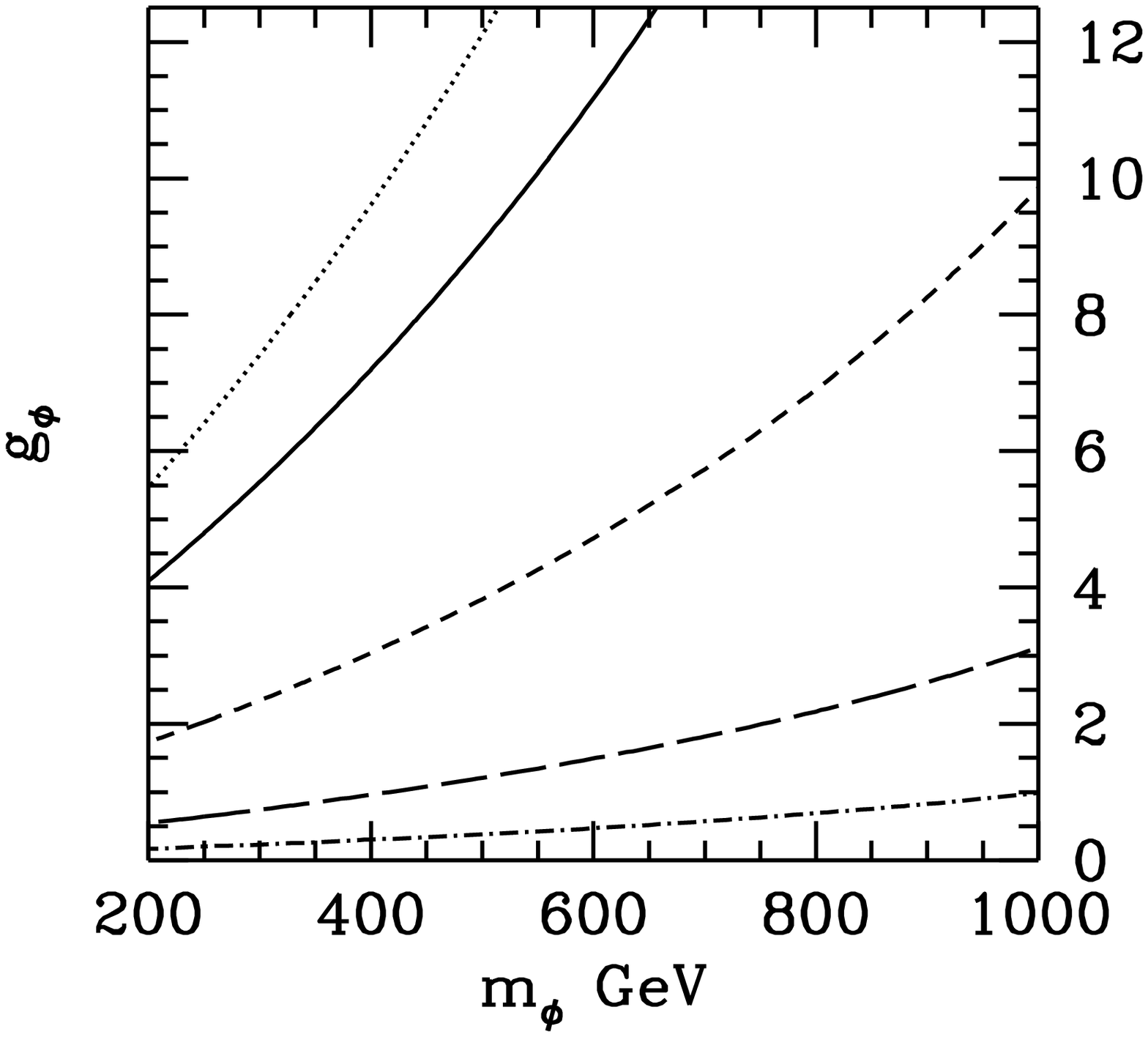}
\caption[dummy]{\small Contours in the mass-coupling plane for different values of
the real parts of EDFF (left) and WDFF (right) of $\tau$ for the $R_2$ model.
Type of lines and corresponding DFF values are the same as
those in Fig.~\ref{fig:conttauimg}.
The solid curve corresponds to the experimental bounds on the dipole
momenta, which are $d_\tau^\gamma=5\times10^{-17}\:e$ cm, 
${\rm Re}\,d_\tau^Z=5.6\times10^{-18}\:e$ cm. EDFF values are at 
a c.m. energy of 4 GeV while WDFF values are at 91.18 GeV. 
Top-quark mass is taken to be 180 GeV in all cases. 
}
\label{fig:conttaureal}
\end{figure}

At a fixed c.m. energy the form factors are functions of two parameters --
the mass and the coupling of the leptoquark considered. From an assumed value
of the form factor it is possible to get contours in the plane of the 
mass and coupling constant
of the leptoquark. Contours in the mass-coupling plane for the doublet
$R_2$ leptoquark model are given for different values of tau lepton DFF's 
in Figs.~\ref{fig:conttauimg} and  \ref{fig:conttaureal}. 
Validity of perturbation theory allows 
values of coupling $g_\phi<4\,\pi$. In the case of the tau lepton we
have considered the present experimental limits on the dipole moments.

Fig.~\ref{fig:conttaureal} shows
the allowed region in the $m_\phi$-$g_\phi$ plane which lies below the solid
line if we consider the present
experimental limit on the  tau electric dipole moment.
No value of $m_{\phi}$ or $g_{\phi}$ is completely excluded. However,
it is possible that 
future experiments may  give more stringent limits on dipole form factors which 
may put upper bounds on the coupling or lower bounds on the mass.

\begin{figure}
\vskip 6cm
\includegraphics{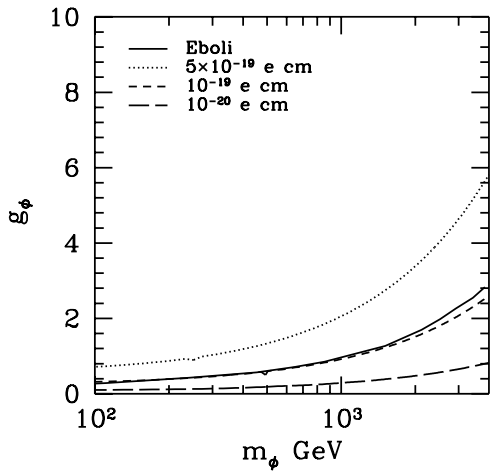}
\caption[dummy]
{\small Contours in the mass-coupling plane for the $R_2$ leptoquark model.
The solid line represents the limits from the one-loop
calculation by \' Eboli \cite{Eboli} while the other curves correspond to 
different values of the real part of the 
tau EDFF. Top-quark mass of 175 GeV is assumed.
The c.m. energy used is 500 GeV.}
\label{fig:conteboli}
\end{figure}

As mentioned before, LEP results on $Z$ partial widths have been used
to obtain constraints on masses and couplings for third-generation 
leptoquarks \cite{Bhatta,Mizui,Eboli}.  We have chosen \' Eboli's \cite{Eboli}
limits to compare with the constraints we get from the dipole form factors. The
other limits would give similar results.
Fig.~\ref{fig:conteboli} shows contours for different values of 
the real  part of the electric dipole form
factor of $\tau$ along with the limit obtained by \' Eboli \cite{Eboli} in the 
$m_{\phi}$-$g_{\phi}$ plane.
To accommodate their limits we have to restrict the electric dipole form
factors of $\tau$ to be smaller than about $10^{-19}\;e\,{\rm cm}$. A similar
analysis shows that the weak
dipole form factors must be smaller than about $10^{-20}\;e\,{\rm cm}$.

The best limits on $m_\phi$ and $g_\phi$ obtainable from the 
experimental limits on form factors is that from the real part of the weak
dipole moment of the tau lepton, and we use that limit in what follows.  

\begin{figure}
\vskip 5.5cm
\includegraphics{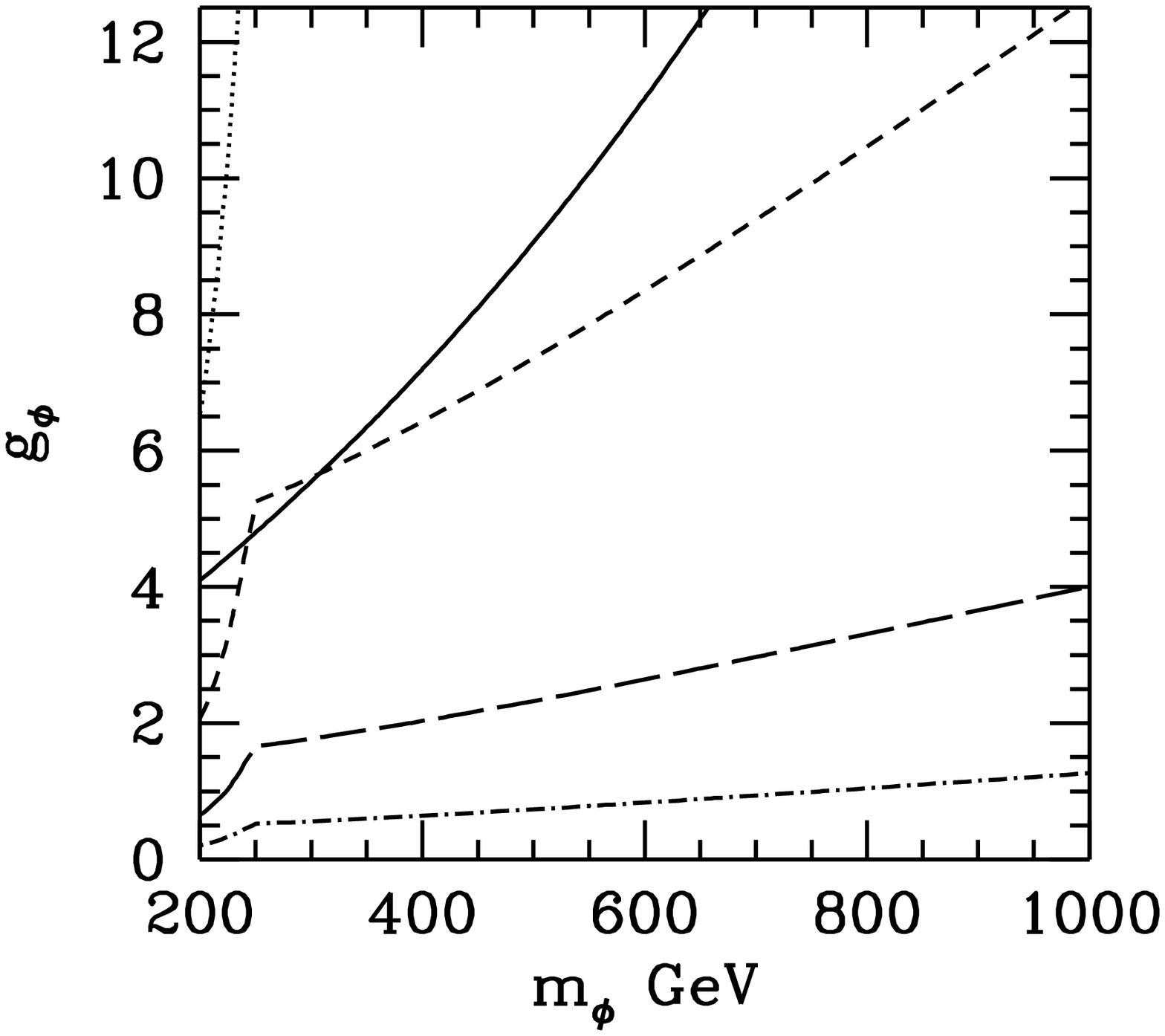}
\includegraphics{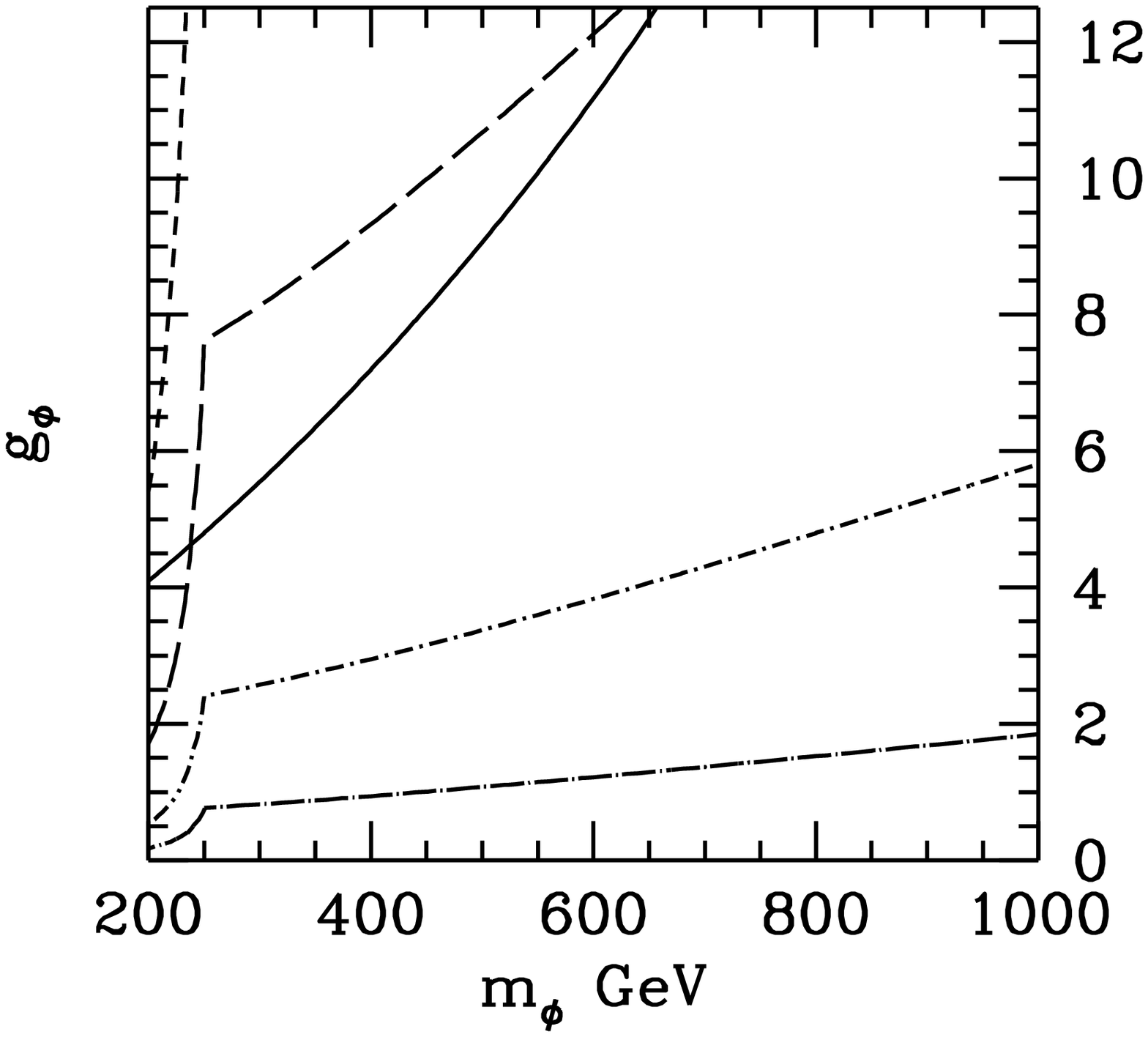}
\caption[dummy]
{\small Contours in the mass-coupling plane for different values of
imaginary parts of the top EDFF (the figure on the left) and WDFF (the figure on the 
right) 
Dotted, dashed, long dashed, dash-dotted and long dash-dotted lines 
correspond to DFF values of
$10^{-18}$, $10^{-19}$, $10^{-20}$, $10^{-21}$ and $10^{-22}\;e$ cm 
respectively.
The solid curve corresponds to the experimental bound on the real part of
the weak dipole moment of $\tau$, $5.6\times10^{-18}\:e$ cm. 
A c.m. energy of 500 GeV and a top-quark mass of 180 GeV are assumed.}
\label{fig:conttopimg}
\end{figure}

\begin{figure}
\vskip 5.5cm
\includegraphics{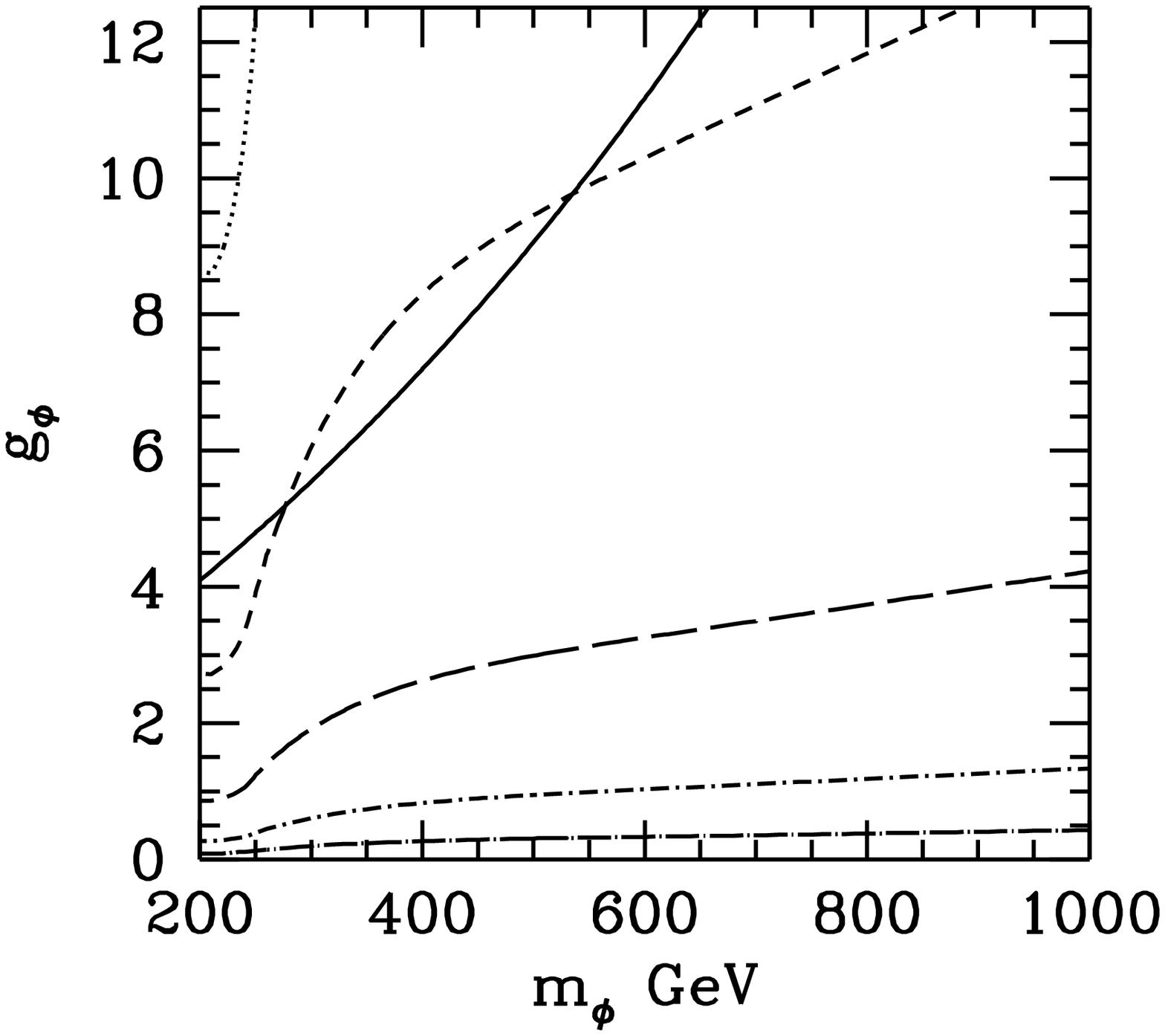}
\includegraphics{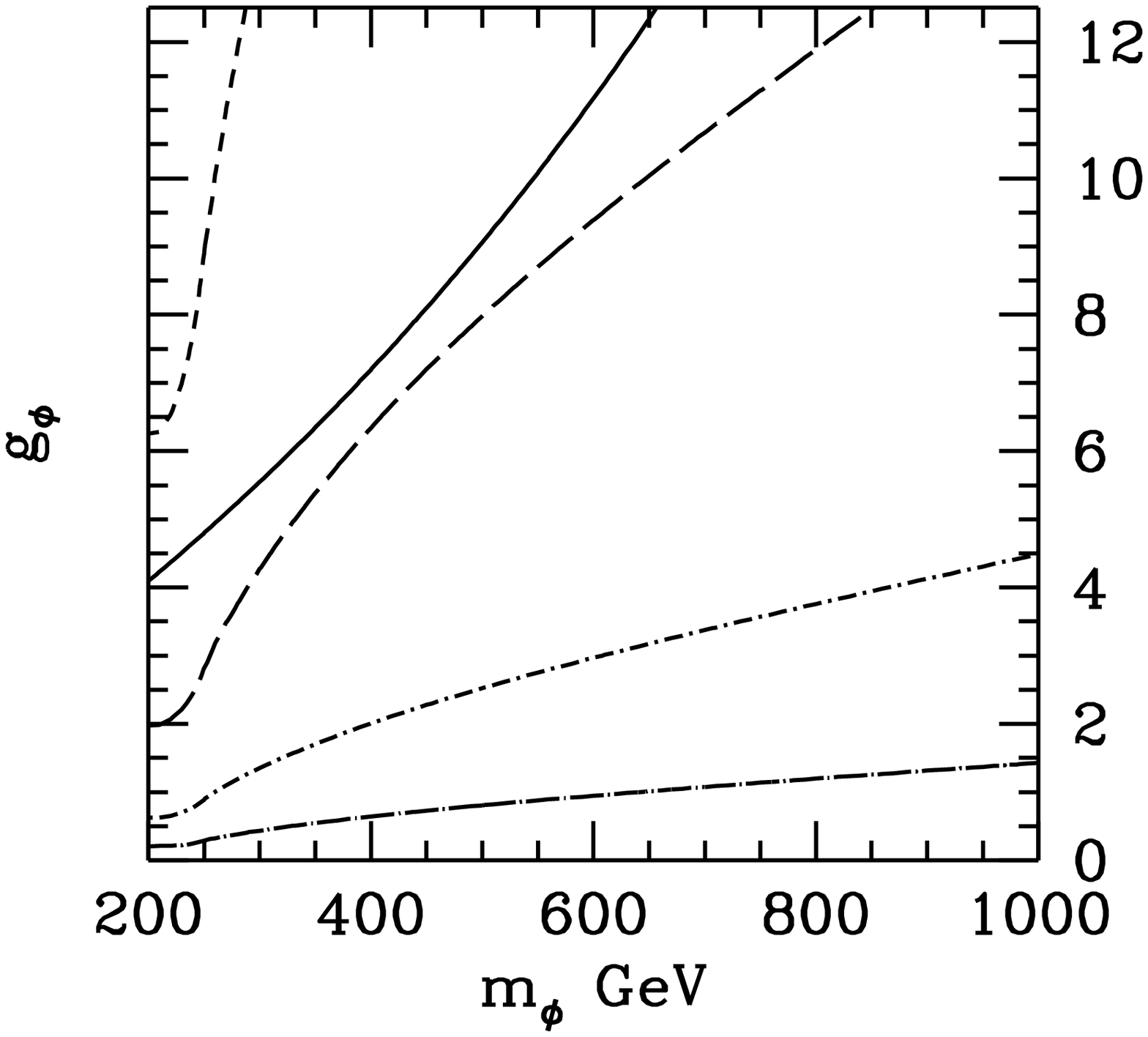}
\caption[dummy]{\small Similar to Fig.~\ref{fig:conttopimg} but with contours 
from the real part
of DFF's. The solid lines are the same as in the Fig.~\ref{fig:conttopimg}.}
\label{fig:conttopreal}
\end{figure}

In the case of top quark there are no experimental limits. From 
the constraints obtained on the mass and the coupling from the 
experimental
bound on the real part of the tau WDFF, Figs.~\ref{fig:conttopimg} and
\ref{fig:conttopreal} show that the top quark
can have values for the imaginary part of the EDFF
as high as $10^{-19}\;e\,{\rm cm}$ except for a 
small mass range ($\sim$ 250-300 GeV). 
The imaginary part of the weak dipole
form factor of the top quark is allowed to be almost as high as
$10^{-20}\;e\,{\rm cm}$.
We get more or less the same limits for the real part of 
both the electric and the weak dipole form factors. 

Again, to
accommodate constraints on the masses and couplings from \' Eboli's result 
we have to have electric dipole
form factors of the order of $10^{-22}\;e$ cm or less and weak 
dipole form factors of the order of $10^{-23}\;e$ cm or less.

\section{Conclusions}

We have analyzed top and tau electric and weak dipole form factors in
models with scalar leptoquarks of weak isospin 0, $\frac{1}{2}$ and 1. We calculate
form factors for different values of $\sqrt{s}$. In general, for leptoquark 
couplings in
the perturbative region and masses allowed by direct experimental searches,
large EDFF and WDFF values for both top and tau are possible. In case of the
top, the values can be as high as $10^{-20}\,e$ cm, whereas tau form factors
can be of the order of $10^{-18}\,e$ cm. The values of form factors are larger
for the case of weak isosopin $\frac{1}{2}$, and we have therefore concentrated 
on that case.

We also obtain contours
in the mass-coupling plane corresponding to fixed values of form factors 
for a given $\sqrt{s}$, and use the experimental limits from LEP on the tau
form factors to obtain the allowed region in the plane. This gives constraints
on the top form factors. For most of the range of couplings and leptoquark mass,
EDFF of the top can be as large as 10$^{-19}$ $e$ cm, and the WDFF can be as 
large as $10^{-20}\,e$ cm. We have also used the indirect constraints on the
mass and coupling of scalar leptoquarks derived from LEP measurements of the
one-loop contribution of leptoquarks to the $Z$ partial decay widths. These
constraints are  more stringent, and do not permit tau EDFF above about 
$10^{-19} \,e$ cm and tau WDFF above about $10^{-20}\,e$ cm. 
The corresponding upper 
bounds on the top EDFF and WDFF are, respectively, $10^{-22}\,e$ cm and 
$10^{-23}\,e$ cm.

We thus conclude that though large dipole form factors are allowed in the 
scalar 
leptoquark model with parameter values consistent with direct experimental
constraints, the indirect constraints from one-loop contributions to $Z$
decay parameters seem to permit only values of form factors which lie 
below the range likely to be explored in experiments in the foreseeable future.

It would be interesting to investigate the one-loop contribution of the 
third-generation
leptoquarks to CP violation in the decay $t \rightarrow b W$.

Finally, we end with a few comments on comparison of our work with other recent work on DFF 
in third-generation leptoquark
models \cite{mahanta,bernlq}.  Mahanta \cite{mahanta} has estimated the
$\tau$ EDFF and has reached the conclusion that it can be as high as
$10^{-19}\;e\,{\rm cm}$ for a choice of $g_\phi$ amd $m_\phi$ consistent with
experimental constraints.  He does not discuss the momentum dependence of
DFF's.  Bernreuther {\it et al.} \cite{bernlq} have obtained 
the $\sqrt{s}$ dependence of the DFF's of $\tau$.  Our results are in
agreement with theirs when an overall scale error in their curves is taken into
account \cite{private}.
Neither of \cite{mahanta} and \cite{bernlq} discuss DFF's of both $t$
and $\tau$ in a concerted manner as we have done here.

\vskip 1cm
\noindent {\Large \bf
Acknowledgements}
\vskip .5cm

We acknowledge with thanks the 
initial collaboration of Torsten Arens. We also thank 
Frank Cuypers for lending his contour-plotting program.
One of us (S.D.R.) thanks Prof. J.W.F. Valle for his warm hospitality at 
the University of Valencia where he spent a sabbatical year. 
This work was supported by DGICYT under grant PB95-1077 and by the TMR
network grant ERBFMRXCT960090 of the European Union. S. D. R. was
supported by DGICYT sabbatical grant SAB95-0175.

\end{document}